\documentclass[a4paper,11pt,oneside]{article}
\usepackage{xspace}
\usepackage[table]{xcolor}
\usepackage[colorlinks=true,urlcolor=blue,linkcolor=blue,citecolor=blue]{hyperref}  % for DOI link
\usepackage[margin=1cm,font=small,labelfont=bf, labelsep=period]{caption}
\usepackage[sf,bf,compact,topmarks,calcwidth,pagestyles]{titlesec}

\usepackage{amsmath}
\usepackage{amssymb}
\usepackage{graphicx}
\usepackage{geometry}
\usepackage{url}

\title{A Novel Hadronic Calorimeter With A Direct Neutron Readout}
\author{I. Giomataris, F. Jeanneau, T. Papaevangelou, G. Tsiledakis\thanks{Email: \texttt{tsiledak@gmail.com}}, M. Vandenbroucke}
\date{\vspace{+0.5 mm} \textit{CEA Irfu, University Paris-Saclay, F-91191 Gif-sur-Yvette, France} \\
\vspace{3 mm} November 26, 2026}
\begin{document}

\maketitle

\begin{abstract}

A neutron-sensitive sampling calorimeter based on alternating lead absorber and gadolinium-loaded liquid scintillator layers is investigated using detailed Geant4 Monte Carlo simulations. In addition to the conventional prompt calorimetric signal, the proposed detector records delayed energy from neutron moderation and capture, providing direct information on the neutron component of hadronic showers.

A six-layer Pb/LAB-Gd calorimeter exposed to 10~GeV protons is studied to characterize its neutron response and evaluate its impact on calorimetric performance. The delayed deposited energy is found to be almost perfectly proportional to the neutron-capture multiplicity, providing a direct calibration of the neutron-sensitive signal. Event-by-event analyses further reveal a clear relationship between the prompt calorimetric response and the delayed neutron observable, demonstrating that the latter contains substantial information on the invisible hadronic energy.

Exploiting this correlation through a nonlinear event-by-event
correction based on branch-wise interpolation reduces the prompt-energy
resolution from 21.8\% to 13.3\% for the simulated sample, without
rejecting events. This result is obtained using the same event sample
to construct and evaluate the correction and should therefore be
regarded as preliminary until validated on an independent sample.
Furthermore, the analysis of events with similar neutron multiplicities
indicates that fluctuations in neutron production are an important
contribution to the overall variation in the prompt calorimetric
response. These results demonstrate the potential of gadolinium-loaded
sampling calorimeters to provide additional information on hadronic
showers and motivate further studies with independent validation and
more realistic calorimeter configurations.

\end{abstract}

\section{Introduction}

Hadronic calorimeters (HCALs) are essential detector systems in high-energy physics experiments~\cite{PDG2024} for the measurement of the energy of strongly interacting particles such as protons, neutrons, charged pions, and hadronic jets. When an energetic hadron enters dense matter, it initiates a hadronic shower through successive nuclear interactions, producing large numbers of secondary particles including charged hadrons, photons, nuclear fragments, and neutrons.

The characteristic scale governing the longitudinal development of a hadronic shower is the nuclear interaction length, $\lambda_I$. In order to contain most of the shower energy, hadronic calorimeters typically require total thicknesses of approximately 5--8 interaction lengths. Because of the large amount of dense material involved, most HCAL systems employ a sampling structure consisting of alternating passive absorber layers and active detector elements capable of measuring the energy deposited by charged particles.

Only a fraction of the total hadronic shower energy contributes directly to the prompt detector signal. A significant component remains ``invisible'' due to nuclear excitation, nuclear breakup processes and neutron production~\cite{wigmans2000,fabjan2003}. In particular, neutrons carry away an important fraction of the hadronic shower energy and are only weakly detected in conventional calorimeters. Fluctuations associated with this invisible energy represent one of the principal limitations on hadronic energy resolution and motivate the development of detector concepts with enhanced neutron sensitivity~\cite{wigmans2000,akchurin2005}.

In this work we investigate a neutron-sensitive sampling calorimeter concept in which gadolinium-loaded liquid scintillator layers are integrated directly into the calorimeter structure. The proposed detector consists of alternating lead absorber plates and active scintillator layers extending over several nuclear interaction lengths. Each active layer provides two complementary observables:

\begin{itemize}
\item a prompt scintillation signal produced by charged particles crossing the detector,
\item a delayed scintillation signal generated by neutron moderation and subsequent neutron capture.
\end{itemize}

The delayed component provides direct sensitivity to neutrons produced during the hadronic shower and therefore offers an additional observable that may complement the conventional calorimetric measurement.

\begin{figure}[h]
\centering
\includegraphics[width=0.72\textwidth]{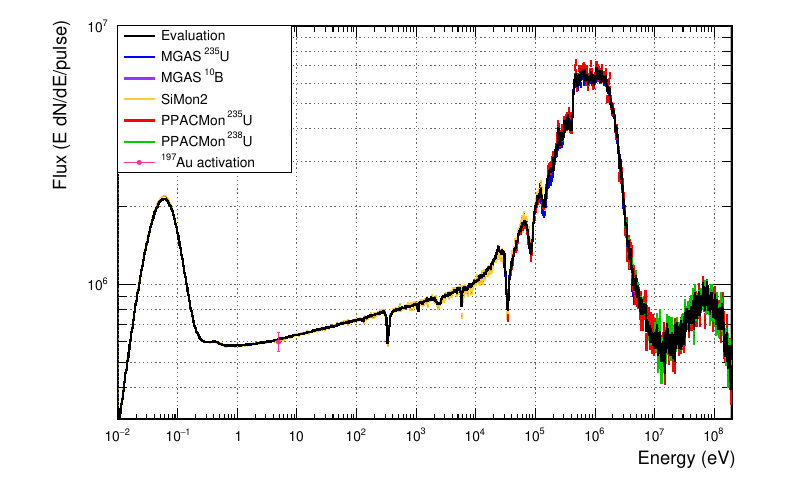}
\caption{Evaluated neutron flux at the CERN n\_TOF EAR2 facility produced by high-energy protons impinging on a thick lead spallation target, adapted from Ref.~\cite{ntof2025}. The neutron spectrum extends over many orders of magnitude in energy, from thermal neutrons up to the GeV region, illustrating the broad neutron component generated inside dense absorber materials.}
\label{fig:ntof_spectrum}
\end{figure}

Hadronic showers produce large neutron multiplicities through spallation and evaporation processes, with neutron yields reaching several tens of neutrons per GeV of incident hadron energy~\cite{wigmans2000}. The neutron energy spectrum extends over many orders of magnitude, from fast neutrons immediately after production down to thermal energies following moderation inside hydrogen-rich materials. Figure~\ref{fig:ntof_spectrum} illustrates a representative neutron spectrum measured at the CERN n\_TOF facility, emphasizing the importance of efficient neutron moderation and thermal-neutron capture in neutron-sensitive calorimeters.

\begin{figure}[h]
\centering
\includegraphics[width=0.75\textwidth]{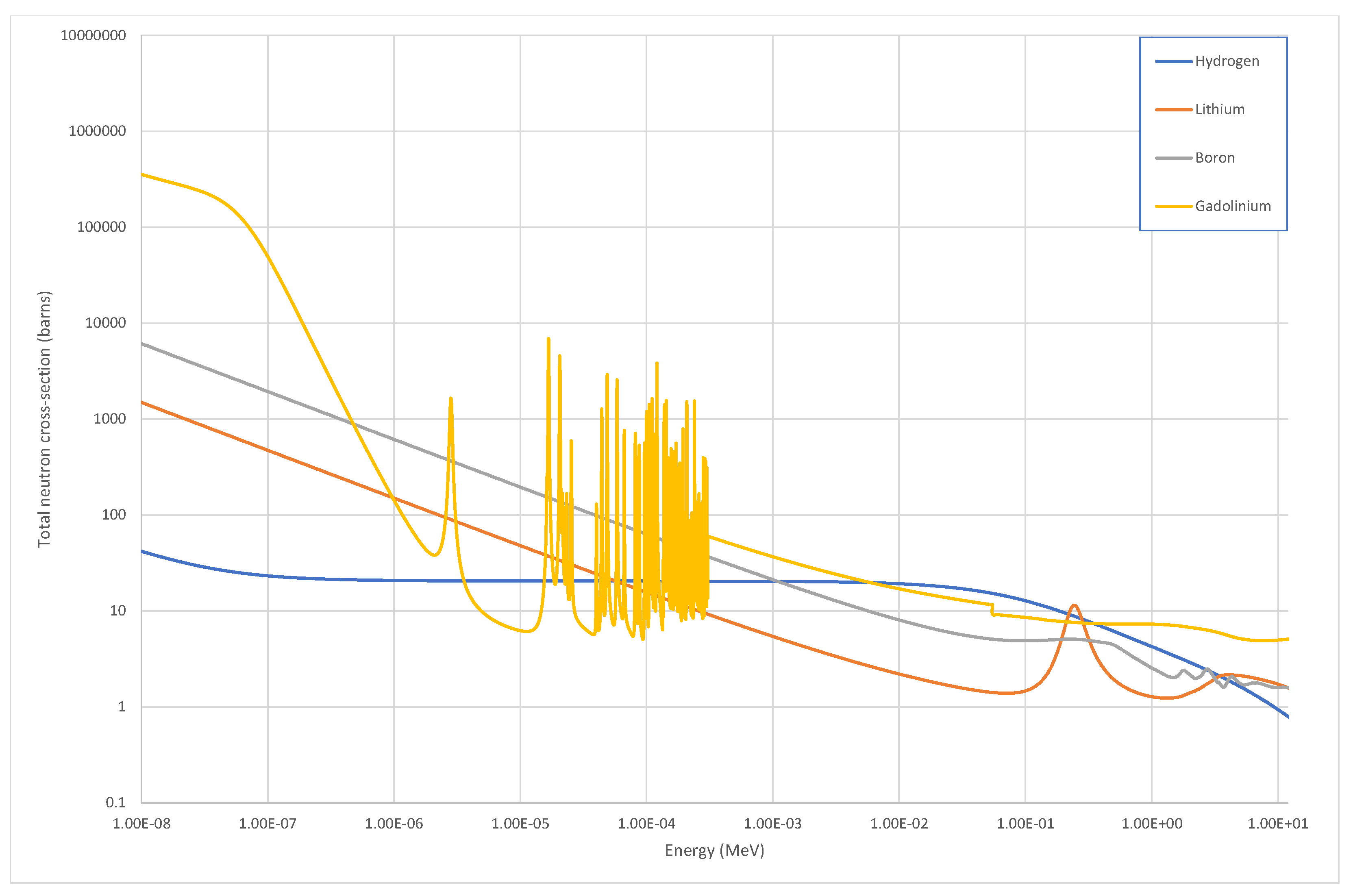}
\caption{Thermal neutron capture cross sections for several neutron-sensitive isotopes commonly used in detector applications, adapted from Ref.~\cite{cieslak2019}. Gadolinium exhibits an exceptionally large capture cross section at thermal energies, exceeding that of hydrogen by several orders of magnitude.}
\label{fig:capture_xsec}
\end{figure}

To detect these neutrons efficiently, liquid scintillators can be loaded with isotopes possessing very large thermal-neutron capture cross sections, including $^{10}$B, $^{6}$Li and gadolinium~\cite{RAleksan}. The use of thermal or quasi-thermal neutron capture as an independent observable of the neutron component of hadronic shower energy was already briefly mentioned in~\cite{RAleksan}. In this approach, the delayed signal associated with neutron capture provides information on the population of low-energy neutrons, offering a potential means of investigating fluctuations in the hadronic calorimetric response. Gadolinium is particularly attractive because of its exceptionally large thermal-neutron capture cross section, of the order of $5\times10^{4}$~barns, and the release of approximately $8$~MeV in a cascade of $\gamma$ rays following neutron capture~\cite{RAleksan}. These characteristics make Gd-loaded scintillators promising for detecting delayed neutron-capture signals, and gadolinium has also been successfully employed in large-scale neutrino detectors~\cite{beacom2004,an2016,Daya2}. A practical consideration, however, is the relatively long characteristic capture time, which can be of the order of $100~\mu\mathrm{s}$ depending on the scintillator composition and geometry, potentially complicating measurements in environments with high event rates or substantial backgrounds~\cite{RAleksan}. Representative thermal-neutron capture cross sections for the relevant isotopes are shown in Figure~\ref{fig:capture_xsec}.

Whereas large neutrino detectors typically employ gadolinium concentrations of approximately 0.1\% by mass in order to preserve long optical attenuation lengths, the much smaller dimensions of calorimeter cells considerably relax these constraints. Consequently, higher gadolinium concentrations become feasible and may significantly improve neutron-capture efficiency while simultaneously reducing neutron-capture times.

As scintillation medium we consider linear alkylbenzene (LAB), which combines good optical properties, chemical stability, radiation tolerance and compatibility with gadolinium loading. LAB-based scintillators have been successfully employed in several modern neutrino experiments~\cite{Daya2,dayaBayLS}.

Whereas established homogeneous electromagnetic calorimeters often
rely on dense inorganic scintillating crystals, such as PbWO$_4$,
CsI and BGO~\cite{Shwartz2002,Lecoq1995}, the proposed approach
employs a hydrogen-rich, Gd-loaded liquid scintillator as the
active medium. An important feature of this choice is the dual role of the scintillator as both an active calorimetric medium and a neutron-sensitive material. Owing to its high hydrogen content, LAB provides an efficient medium for the moderation of fast neutrons through elastic scattering, while the gadolinium loading enhances their subsequent capture. The energy released in the capture $\gamma$-cascade is then detected through the scintillation response of the same active medium, providing a delayed signal in addition to the prompt scintillation generated by the hadronic shower. This combination offers an additional observable of the neutron component of the shower, which can potentially be exploited to investigate and correct fluctuations in the calorimetric response.

For photodetection, avalanche photodiodes (APDs) are considered
a possible readout option because of their high quantum efficiency,
large dynamic range and demonstrated radiation tolerance in
high-energy physics experiments~\cite{cms_ecal_apd,apd_hamamatsu}. Their nearly linear
response over a wide dynamic range makes them potentially suitable
for measuring both the prompt scintillation signal and the delayed
neutron-capture signal. A detailed assessment of the optical
collection efficiency and photodetector response, however, is
beyond the scope of the present simulation study.

As a possible starting point for an experimental proof-of-principle
demonstrator, we consider cubic scintillator cells with dimensions
of $5\times5\times5~\mathrm{cm}^3$, filled with LAB-based scintillator
and instrumented with photosensors. A reflective internal
surface could be used to improve light collection. The choice
of photosensor, optical configuration and sensor coverage will
require dedicated optical simulations and experimental
characterization. These aspects are therefore considered
design considerations for a future demonstrator rather than
results of the present Geant4 study.

%The proposed detector concept is illustrated in Figure~\ref{fig:concept}. 
Unlike previous neutron detectors positioned downstream of calorimeters for particle identification~\cite{neucal}, the present approach integrates neutron-sensitive active layers directly inside the calorimeter sampling structure. Using Geant4 Monte Carlo simulations, we investigate neutron production, transport, moderation, capture and delayed energy deposition in a multilayer Pb/LAB-Gd calorimeter exposed to 10~GeV protons. Particular emphasis is placed on the influence of gadolinium concentration, neutron-capture multiplicities, delayed energy deposition and the intrinsic calibration of the neutron-sensitive response. The objective of the present work is to establish the feasibility and characterize the performance of neutron-sensitive sampling layers as a first step toward future studies of full calorimetric systems.

\section{Simulation Setup}

The detector response was investigated using Geant4 version 11.2.2~\cite{geant4,geant4update2006,geant42016} with a custom physics list based on the \texttt{FTFP\_BERT\_HP} hadronic physics configuration. 

A monoenergetic 10~GeV proton beam impinges on a lead block of dimensions
$100\times100\times25~\mathrm{cm^3}$,
representing the absorber element of a sampling calorimeter module. Secondary particles produced in the hadronic cascade, including evaporation and spallation neutrons, subsequently enter an active scintillator volume located immediately downstream of the lead target.

The active detector has identical transverse dimensions and a thickness of
25~cm and is filled with linear alkylbenzene (LAB). Since commercial LAB is a mixture of alkylbenzene molecules with different alkyl-chain lengths, it was modeled using the representative molecular composition C$_{18}$H$_{30}$ and a density of 0.86 g cm$^{-3}$. This approximation reproduces the average elemental composition and bulk properties of commercial LAB and is commonly adopted in Monte Carlo simulations of liquid scintillators~\cite{Yeh2011,JUNOCDR,SasolLAB,SNOplus}.

Three detector configurations were investigated:

\begin{itemize}
\item pure LAB,
\item LAB loaded with 0.5 \% gadolinium,
\item LAB loaded with 2 \% gadolinium.
\end{itemize}

The gadolinium concentration is specified by mass fraction. Gadolinium loading preserves the hydrogen content responsible for neutron moderation while introducing an extremely large thermal-neutron capture cross section, resulting in efficient neutron capture accompanied by high-energy capture-$\gamma$ cascades.

During the development of the simulation, complementary calculations were performed using FLUKA~\cite{fluka2025,ballarini2024} to provide an independent cross-check of neutron production and detector response. The two codes predict different absolute neutron multiplicities, with Geant4 yielding approximately 20--25\% higher values for neutrons leaving the Pb absorber and entering the active detector. Nevertheless, both codes reproduce the same qualitative behaviour with increasing gadolinium concentration: neutron leakage from the detector decreases, while the number of captures per incident proton increases. These differences in absolute yields may reflect differences in the underlying physics models and nuclear data. Geant4 was used for the detailed detector study, while FLUKA provided an independent cross-check of the overall neutron yields and their dependence on gadolinium concentration. Unless explicitly stated otherwise, all quantitative results presented in this work are obtained with Geant4.

Neutron interactions below 20~MeV were treated using the Geant4 High Precision (HP) neutron package together with the G4NDL 4.7.1 evaluated nuclear-data library. Thermal-neutron capture cross sections were therefore taken from the G4NDL 4.7.1 capture database, providing isotope-dependent neutron transport and capture for hydrogen and naturally occurring gadolinium.

Except where otherwise noted, all results correspond to analogue Monte Carlo simulations of $10^4$ incident protons.

\section{Results}

\subsection{Neutron Transport and Capture}

Neutron transport from the lead absorber into the active scintillator was investigated using Geant4 (version 11.2.2) with the High Precision (HP) neutron package and the G4NDL 4.7.1 evaluated nuclear-data library. Thermal neutron capture cross sections were taken from the G4NDL4.7.1 Capture library, allowing detailed event-by-event tracking of neutron moderation and capture.

For every primary proton, all neutrons produced inside the Pb absorber were individually followed until either capture or escape from the detector volume. For each capture, the simulation records the capture position, capture time, capture isotope and the total energy carried by the emitted capture-$\gamma$ cascade.

\begin{table}[htbp]
\centering
\caption{Summary of neutron transport and capture observables obtained from the Geant4 simulations. All quantities are given per incident 10 GeV proton.}
\label{tab:transport}

\begin{tabular}{lccc}
\hline
Observable & LAB & LAB+0.5\% Gd & LAB+2\% Gd\\
\hline

Neutrons leaving Pb
&150.80
&150.76
&150.52\\

Neutrons entering detector
&76.35
&76.23
&76.14\\

Neutrons leaving detector
&7.69
&6.19
&6.16\\

Captured neutrons
&49.86
&62.18
&62.85\\

Capture on Gd (\%)
&--
&96.09
&97.63\\

Capture on H (\%)
&98.85
&3.81
&2.10\\

Mean capture time ($\mu$s)
&267.3
&58.9
&35.0\\

Deposited capture energy (MeV)
&439.1
&584.3
&598.7\\

\hline
\end{tabular}
\end{table}

The neutron yield produced inside the lead absorber is practically independent of the scintillator composition, remaining close to 151 neutrons per incident proton for all detector configurations. Likewise, approximately 76 neutrons per proton enter the downstream scintillator volume, confirming that the absorber geometry determines the neutron production while the detector composition primarily affects the subsequent neutron transport and capture.

A clear improvement is observed once gadolinium is introduced into the scintillator. The number of neutrons escaping from the detector decreases from 7.69 neutrons per proton for pure LAB to about 6.2 neutrons per proton for both Gd-loaded configurations, while the capture probability increases from 49.9 to more than 62 captures per proton. Increasing the gadolinium concentration from 0.5\% to 2\% produces only a modest additional increase in the total number of captures, indicating that the neutron-capture efficiency is already close to saturation for the present detector geometry.

The isotope responsible for neutron capture changes dramatically after gadolinium loading. In pure LAB essentially all captures occur on hydrogen, whereas after adding only 0.5\% Gd more than 96\% of all captures occur on gadolinium. Increasing the concentration to 2\% further raises this fraction to almost 98\%, leaving only a very small contribution from hydrogen captures. This transition is a direct consequence of the extremely large thermal-neutron capture cross sections of the naturally abundant isotopes $^{155}$Gd and $^{157}$Gd.

An equally important effect is observed in the capture timing. The mean neutron-capture time decreases from 267~$\mu$s in pure LAB to only 59~$\mu$s for 0.5\% Gd and 35~$\mu$s for 2\% Gd. Such short capture times are particularly attractive for calorimetric applications because they allow the delayed neutron signal to be collected within relatively short acquisition windows while significantly reducing pile-up from captures occurring at very long times.

Finally, the total energy deposited through neutron-capture processes also increases with gadolinium loading, rising from 439 MeV in pure LAB to nearly 600 MeV for the 2\% Gd detector. This increase reflects both the larger number of neutron captures and the substantially higher energy released in the characteristic gadolinium capture-$\gamma$ cascades compared with hydrogen capture.

\subsection{Neutron Capture Timing and Delayed Signal Characteristics}

\begin{figure}[htbp]
\centering
\includegraphics[width=0.75\textwidth]{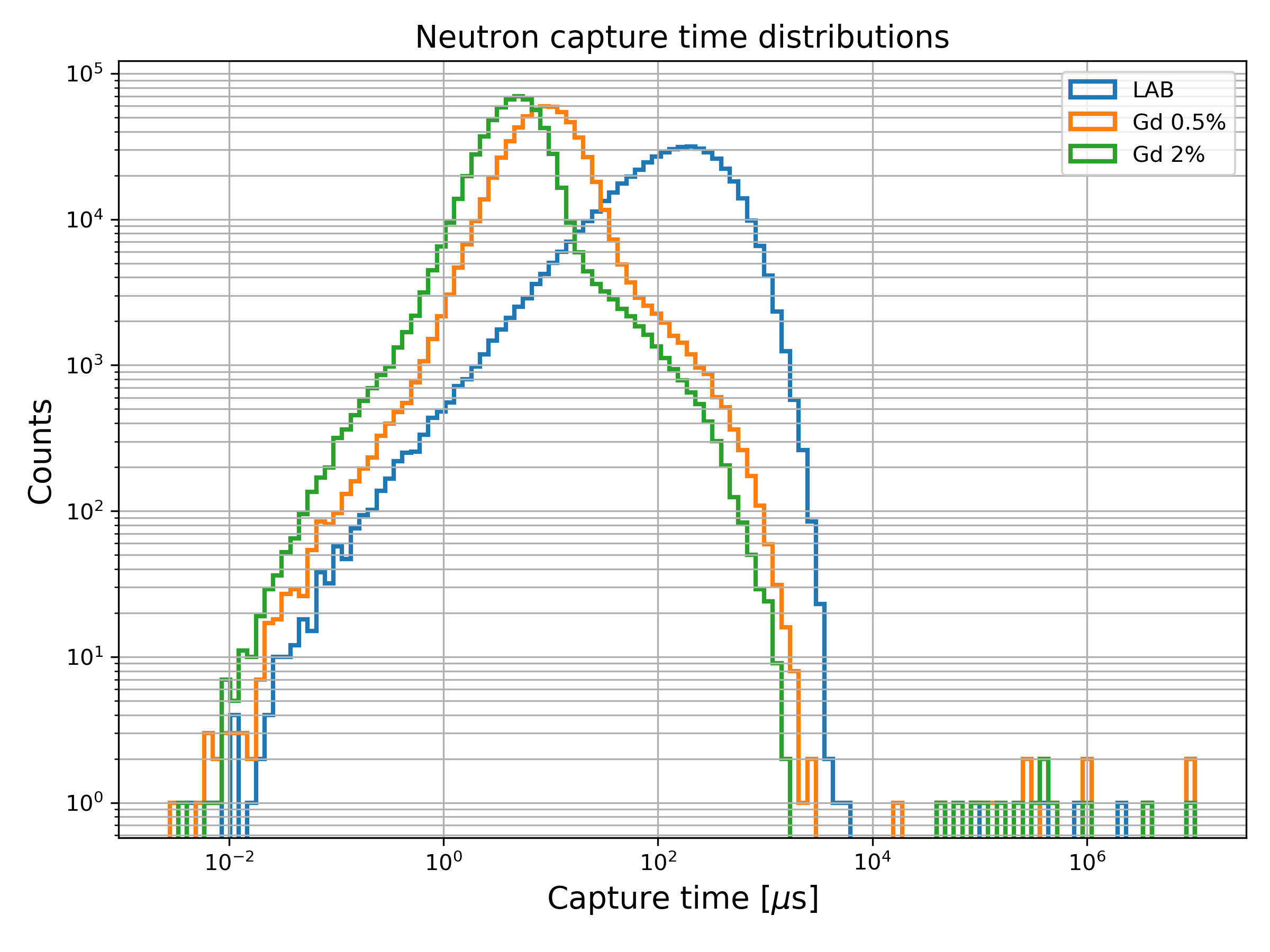}
\caption{
Neutron capture-time distributions for pure LAB, LAB+0.5\% Gd and LAB+2\% Gd obtained with Geant4.
}
\label{fig:capture_time}
\end{figure}

The event-by-event capabilities of Geant4 allow the neutron capture time and the total energy released by the capture-$\gamma$ cascade to be recorded for every neutron captured inside the detector. Figure~\ref{fig:capture_time} compares the capture-time distributions obtained for pure LAB, LAB loaded with 0.5\% Gd and LAB loaded with 2\% Gd.

The addition of gadolinium dramatically accelerates the neutron capture process. While the total number of captured neutrons increases by only about 25\%, the capture-time distribution shifts by almost two orders of magnitude toward shorter times. The median capture time decreases from 135~$\mu$s in pure LAB to only 8.9~$\mu$s for 0.5\% Gd and 4.7~$\mu$s for 2\% Gd, making the delayed neutron signal accessible within acquisition windows compatible with high-rate calorimetry.

The principal capture observables are summarized in Table~\ref{tab:geant_capture_summary}.

\begin{table}[htbp]
\centering
\caption{Characteristic neutron-capture observables obtained from the Geant4 simulations.}
\label{tab:geant_capture_summary}

\begin{tabular}{lccc}
\hline
Material &
Median capture time ($\mu$s) &
Mean capture-$\gamma$ energy (MeV) &
Captures/proton\\
\hline
LAB           &134.8&2.255&49.9\\
LAB+0.5\% Gd  &8.90&7.790&62.2\\
LAB+2\% Gd    &4.70&7.889&62.9\\
\hline
\end{tabular}
\end{table}

\begin{figure}[htbp]
\centering
\includegraphics[width=\textwidth]{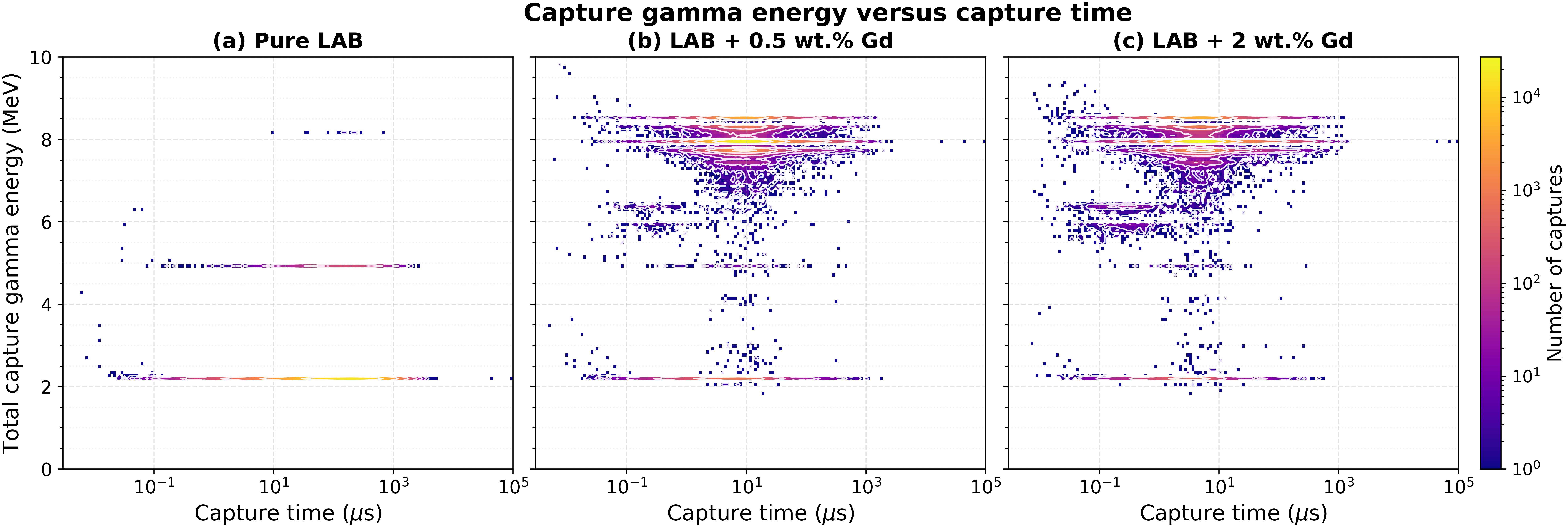}
\caption{
Capture-$\gamma$ energy versus neutron capture time for (a) LAB, (b) LAB+0.5\% Gd and (c) LAB+2\% Gd. The colour scale represents the logarithm of the number of events.
}
\label{fig:capture_gamma_time}
\end{figure}

Besides the faster timing, gadolinium also changes the energy signature of neutron capture. Figure~\ref{fig:capture_gamma_time} shows the total capture-$\gamma$ energy as a function of capture time.

Pure LAB is dominated by neutron capture on hydrogen, producing the characteristic 2.223 MeV capture line. After gadolinium loading, the dominant population shifts to approximately 7.94 MeV, corresponding to the total energy released by the cascade of prompt $\gamma$ rays emitted following neutron capture on $^{155}$Gd and $^{157}$Gd. A smaller population remains at 2.223 MeV due to the residual hydrogen captures, while events between about 5 and 8 MeV originate from partial containment of the multiple capture $\gamma$ rays inside the finite detector volume.

\begin{figure}[htbp]
\centering
\includegraphics[width=0.75\textwidth]{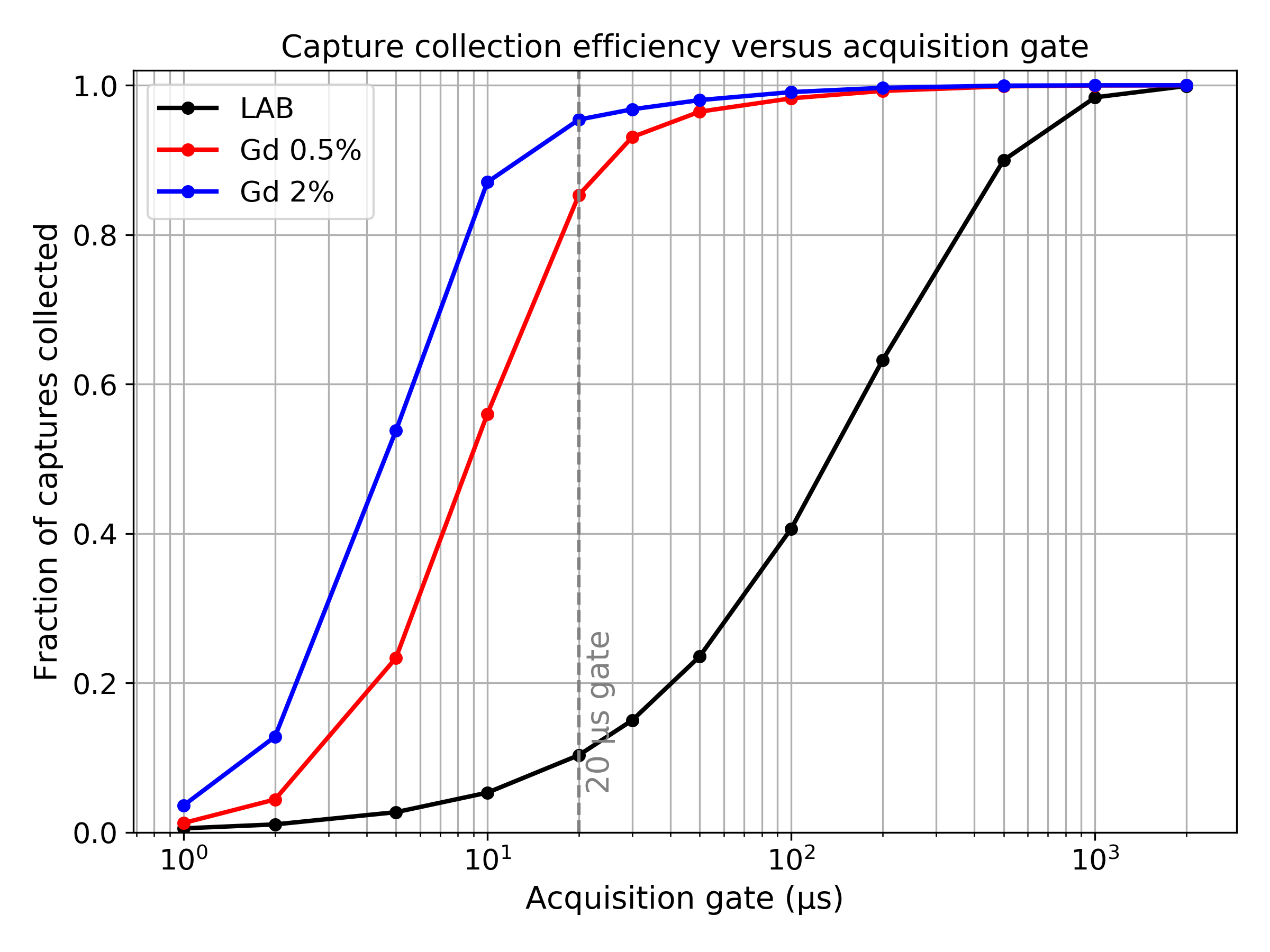}
\caption{
Fraction of neutron captures collected within different acquisition windows.
}
\label{fig:capture_gate_efficiency}
\end{figure}

The practical consequence for calorimetry is illustrated in Fig.~\ref{fig:capture_gate_efficiency}, which shows the cumulative fraction of neutron captures recorded within different acquisition gates.

\begin{table}[htbp]
\centering
\caption{Fraction of neutron captures occurring within several acquisition gates.}
\label{tab:capture_gate_efficiency}

\begin{tabular}{cccc}
\hline
Gate ($\mu$s)
&LAB
&LAB+0.5\% Gd
&LAB+2\% Gd\\
\hline
10  &5.3&56.0&87.0\\
20  &10.3&85.3&95.4\\
30  &15.0&93.1&96.8\\
50  &23.6&96.5&98.0\\
100 &40.7&98.3&99.1\\
\hline
\end{tabular}
\end{table}

The improvement provided by gadolinium is striking. In pure LAB, less than half of all neutron captures occur within 100~$\mu$s, whereas more than 98\% of the captures are already collected within the same acquisition window for both Gd-loaded scintillators. Furthermore, the 2\% Gd detector records nearly 90\% of all neutron captures within only 10~$\mu$s.

These results demonstrate that the principal benefit of gadolinium loading is not a large increase in the total number of neutron captures, but rather the transformation of the delayed neutron signal into a fast, high-energy signature that can be collected within realistic detector integration times. This behaviour is particularly attractive for sampling calorimeters operating in high-rate environments and provides the motivation for extending the study to a realistic multilayer calorimeter geometry in the following section.
%%%%%%%%%%%%%%%%%%%%%%%%%%%%%%%%%%%%%%%%%%%%%%%%%%%%%%%%%%%%%%
%%%%%%%%%%%%%%%%%%%%%%%%%%%%%%%%%%%%%%%%%%%%%%%%%%%%%%%%%%%%%%%%%%%%%%%%%%%%%%%
\subsection{Application to a Multi-Layer Sampling Calorimeter}

The single-module studies presented above establish the fundamental neutron moderation and capture processes occurring inside a Gd-loaded scintillator. In a realistic calorimeter, however, neutrons produced at different stages of the hadronic cascade propagate through several absorber and scintillator layers before being captured. It is therefore important to verify that the delayed neutron signature remains observable once the detector is extended to a sampling calorimeter geometry.
\begin{figure}[htbp]
\centering
\includegraphics[width=0.85\textwidth]{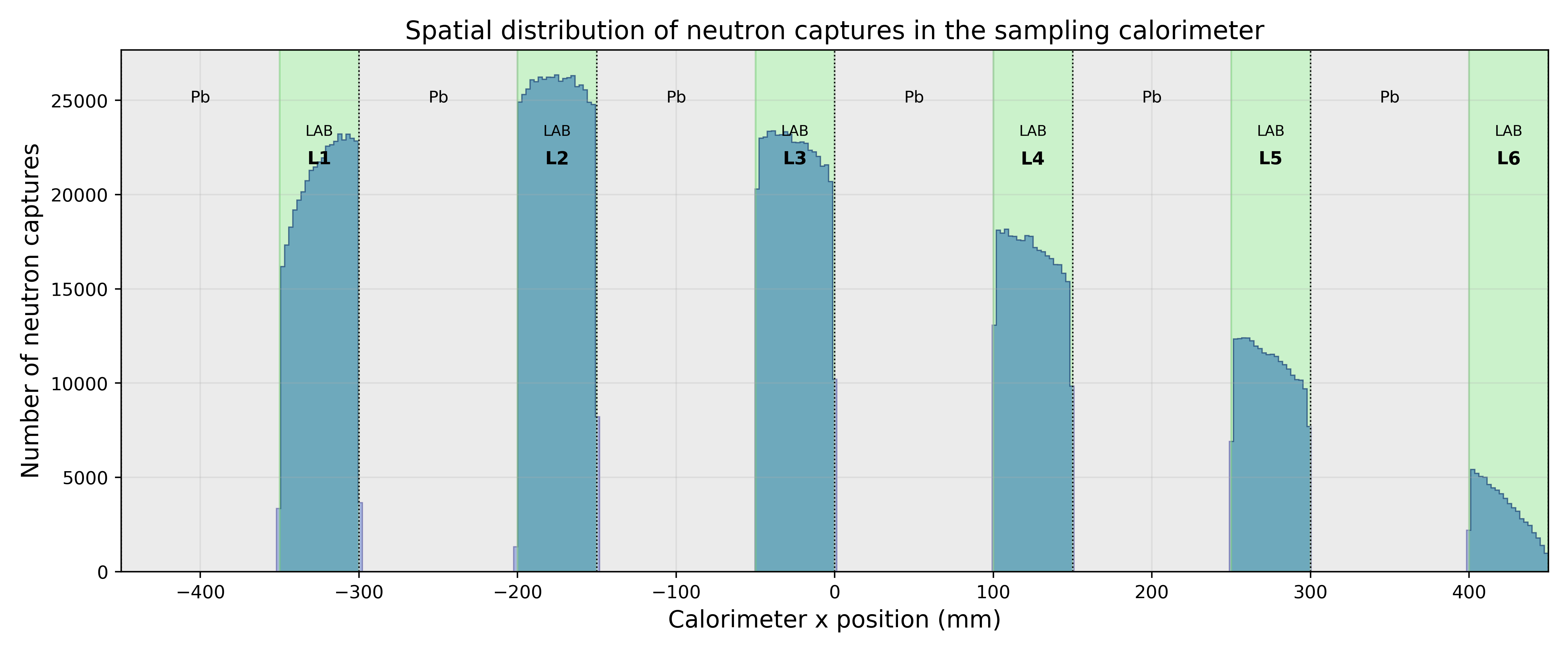}
\caption{
Spatial distribution of neutron captures inside the six-layer Pb/LAB+0.5\% Gd sampling calorimeter obtained with Geant4.
}
\label{fig:layer_capture_map}
\end{figure}

To investigate this behaviour, a prototype calorimeter composed of six sampling units was simulated using Geant4. Each unit consisted of a 10~cm Pb absorber followed by a 5~cm layer of LAB loaded with 0.5\% Gd, resulting in a total detector thickness of 90~cm, corresponding to approximately 107.5 radiation lengths and 3.66 nuclear interaction lengths. The transverse dimensions were $100\times100~\mathrm{cm^2}$ and the detector was irradiated with 10~GeV protons using $10^4$ simulated 10~GeV proton events.

For this geometry, the scintillator layers collected an average deposited energy of approximately 907~MeV per incident proton. About 307 neutrons per proton were produced inside the Pb absorbers, of which nearly 277 entered the scintillator layers. Only about 13 neutrons per proton escaped from the calorimeter, while approximately 197 neutron captures per proton occurred inside the active detector volume. The corresponding median neutron-capture time was 11.1~$\mu$s, demonstrating that the rapid delayed response observed for the single-module configuration is preserved in the complete sampling structure.

\begin{figure}[htbp]
\centering
\includegraphics[width=0.75\textwidth]{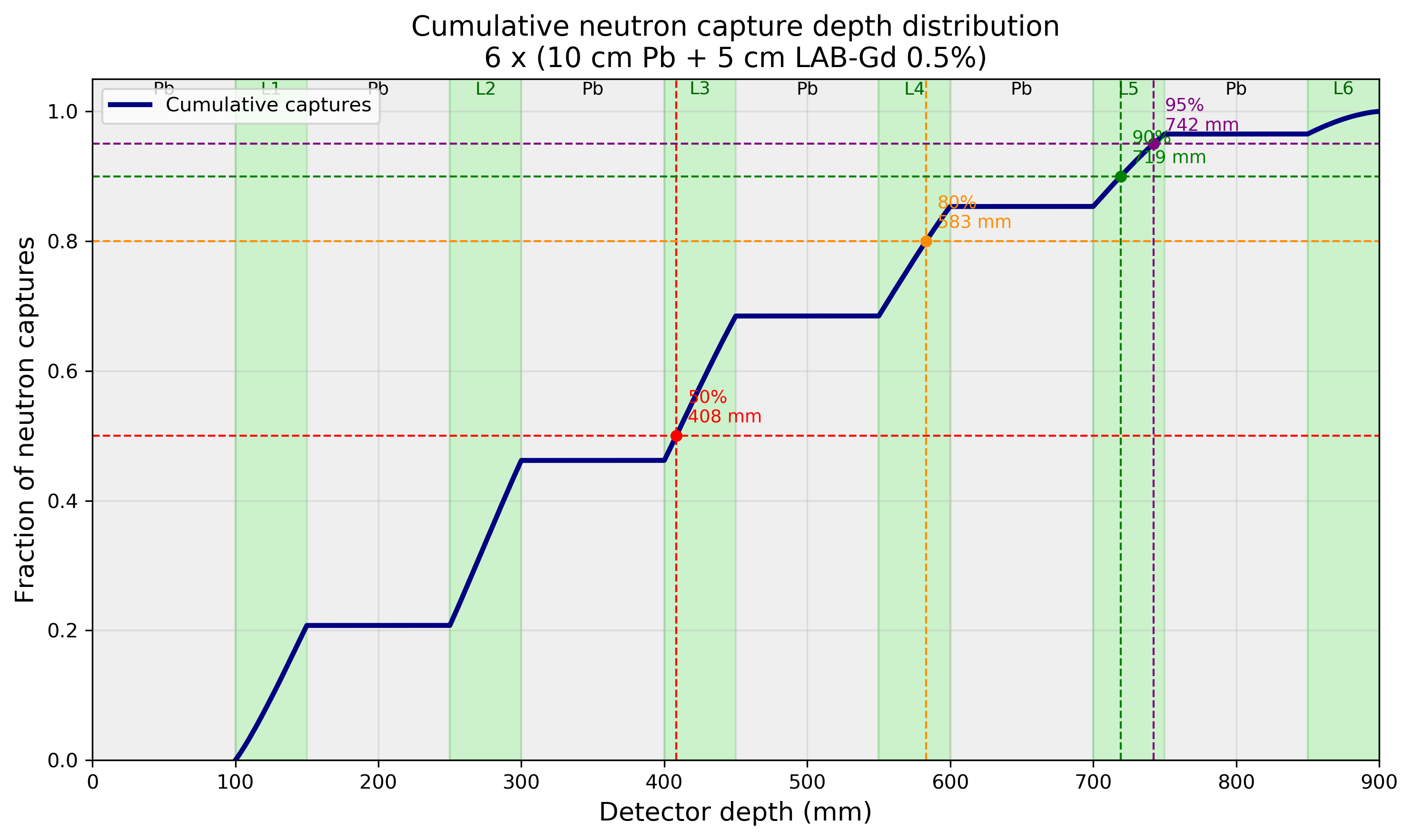}
\caption{
Cumulative neutron-capture depth distribution in the six-layer Pb/LAB+0.5\% Gd calorimeter.
}
\label{fig:capture_depth}
\end{figure}
Figure~\ref{fig:layer_capture_map} shows the spatial distribution of neutron captures throughout the six-layer calorimeter.

The neutron captures are concentrated predominantly in the first half of the detector. The first three scintillator layers account for almost 70\% of all neutron captures, with the second active layer exhibiting the highest capture density. This behaviour reflects the progressive moderation of fast neutrons generated in the upstream Pb absorbers followed by their efficient capture once thermal equilibrium is reached inside the hydrogen-rich scintillator.

The longitudinal evolution of the delayed neutron signal is illustrated in Fig.~\ref{fig:capture_depth}, which presents the cumulative neutron-capture depth distribution measured from the calorimeter entrance.

The cumulative distribution demonstrates that neutron captures remain well localized inside the detector despite the extended moderation process. Approximately 50\% of all captures occur within the first 40~cm of the calorimeter, while more than 90\% are recorded before a depth of about 72~cm. Only a small fraction of captures takes place near the downstream end of the detector.

These observations have important implications for the design of neutron-sensitive sampling calorimeters. First, the delayed neutron signal remains spatially correlated with the hadronic shower development rather than being distributed uniformly throughout the detector. Second, the concentration of neutron captures in the upstream layers indicates that efficient neutron tagging can be achieved without requiring extremely deep calorimeters. Finally, the persistence of neutron captures in the downstream layers demonstrates that delayed neutron information continues to be generated after the prompt shower has largely developed, providing an additional observable that is complementary to the prompt calorimetric response.

The six-layer geometry presented here serves as a proof-of-principle demonstration that the neutron-sensitive concept established for a single detector module naturally extends to a realistic sampling calorimeter. In the following section, the event-by-event response of this prototype calorimeter is analysed in greater detail by investigating the correlation between prompt energy deposition, delayed neutron-capture energy and neutron capture multiplicity.
%%%%%%%%%%%%%%%%%%%%%%%%%%%%%%%%%%%%%%%%%%%%%%%%%%%%%%%%%%%%%%%%%%%%%

%%%%%%%%%%%%%%%%%%%%%%%%%%%%%%%%%%%%%%%%%%%%%%%%%%%%%%%%%%%%%%%%%%%%%%%%%%
\subsection{Performance of the Six-Layer Neutron-Sensitive Calorimeter}
%%%%%%%%%%%%%%%%%%%%%%%%%%%%%%%%%%%%%%%%%%%%%%%%%%%%%%%%%%%%%%%%%%%%%%%%%%

The neutron transport studies presented in the previous section demonstrate that the proposed six-layer calorimeter efficiently moderates and captures evaporation neutrons produced in hadronic showers. In this section, the detector response is investigated on an event-by-event basis using $10^4$ simulated 10~GeV proton events. The objective is to evaluate whether the delayed neutron signal can provide quantitative information on the invisible hadronic energy~\cite{wigmans2000} and consequently improve the calorimetric energy resolution.

The simulations produced, on average, 307.4 neutrons per proton leaving the lead absorbers, of which 276.5 entered the active scintillator layers. Approximately 196.6 neutrons were finally captured inside the detector, corresponding to an average capture efficiency of about 71\% for neutrons entering the active volume with a median neutron-capture time of 11.1~$\mu$s.
%%%%%%%%%%%%%%%%%%%%%%%%%%%%%%%%%%%%%%%%%%%%%%%%%%%%%%%%%%%%%%%%%%%%%%%%%%
\subsubsection{Neutron-sensitive detector response}
%%%%%%%%%%%%%%%%%%%%%%%%%%%%%%%%%%%%%%%%%%%%%%%%%%%%%%%%%%%%%%%%%%%%%%%%%%

\begin{figure}[htbp]
\centering
\includegraphics[width=0.47\textwidth]{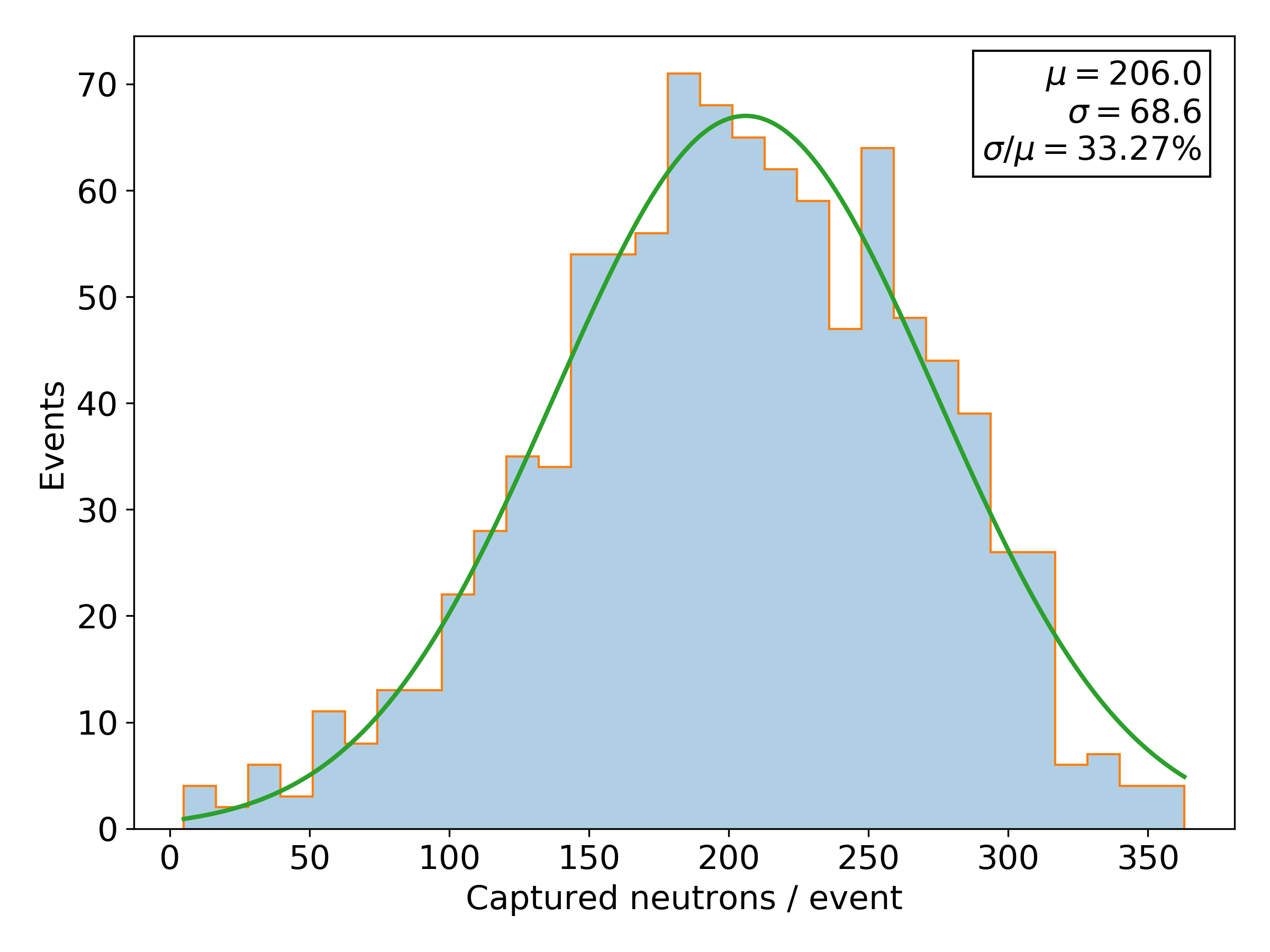}
\hfill
\includegraphics[width=0.47\textwidth]{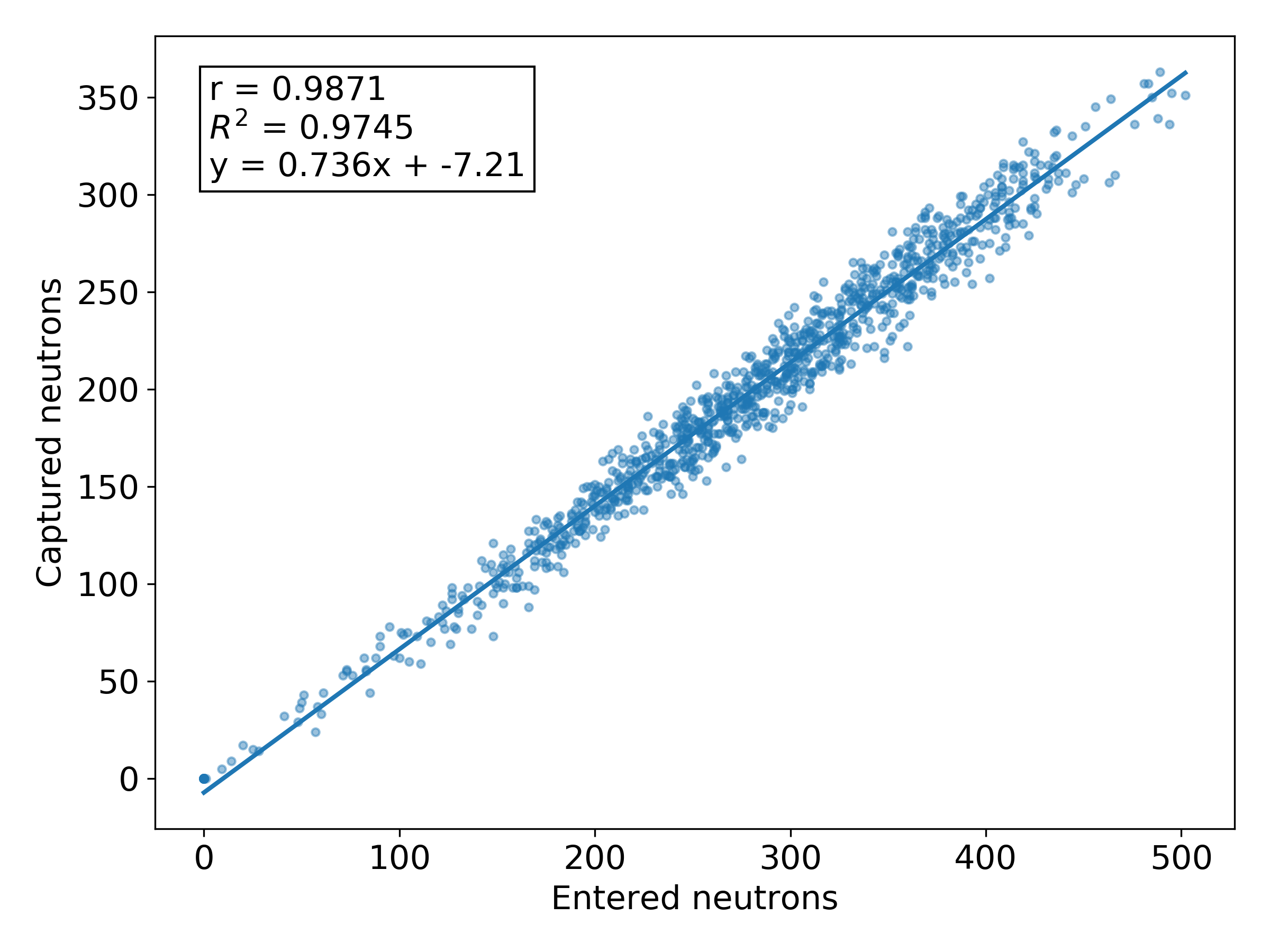}
\caption{
Performance of the neutron-sensitive detector.
(a) Distribution of the captured neutron multiplicity.
(b) Correlation between neutrons entering the active detector and neutrons finally captured.
}
\label{fig:neutronresponse}
\end{figure}

Figure~\ref{fig:neutronresponse} summarizes the basic response of the neutron-sensitive calorimeter.

The event-by-event distribution of the captured-neutron multiplicity
is approximately Gaussian, with a mean value close to 200 captured
neutrons per event. This provides a measure of the statistical
distribution of the delayed neutron signal in the simulated sample.

The correlation between the number of neutrons entering the active
detector and the number finally captured is shown in Fig.~\ref{fig:neutronresponse}b. A strong linear correlation is observed ($R^2=0.975$), indicating that the captured-neutron multiplicity is closely related to the neutron population entering the active scintillator, despite the
stochastic nature of neutron moderation and capture.

This correlation supports the use of the delayed neutron signal as
an observable of the neutron component of hadronic showers. The
event-by-event fluctuations around the correlation are examined
further in the following section.
%%%%%%%%%%%%%%%%%%%%%%%%%%%%%%%%%%%%%%%%%%%%%%%%%%%%%%%%%%%%%%%%%%%%%%%%%%
\subsubsection{Calibration and neutron-counting performance}

Before investigating calorimetric corrections, it is important to
characterize the neutron-counting performance of the proposed detector.
The following measurements examine the relationship between the delayed
signal and the neutron population produced in the hadronic shower, as
well as the correlation between the prompt and delayed calorimetric
responses.

\begin{figure}[htbp]
\centering
\includegraphics[width=0.31\textwidth]{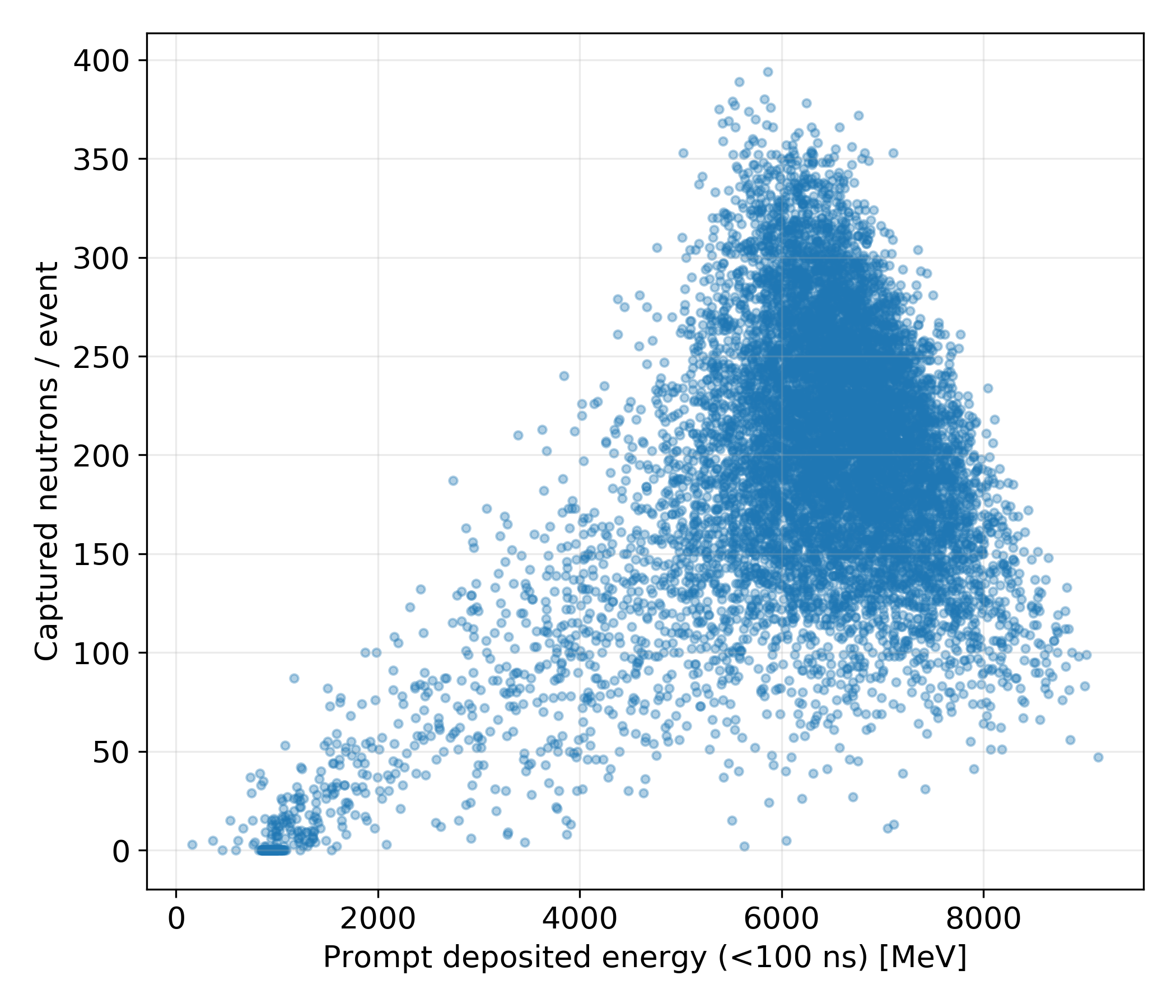}
\hfill
\includegraphics[width=0.31\textwidth]{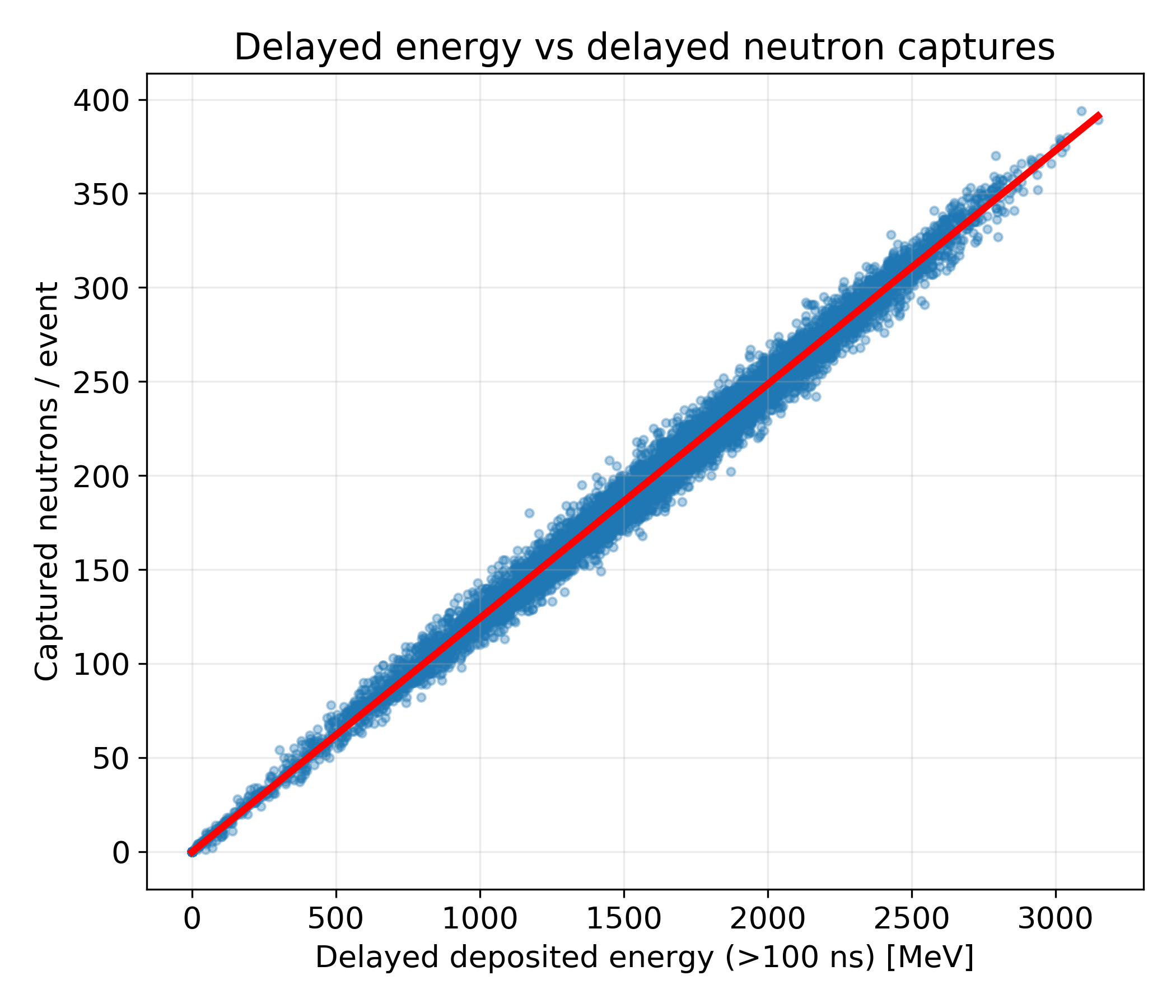}
\hfill
\includegraphics[width=0.31\textwidth]{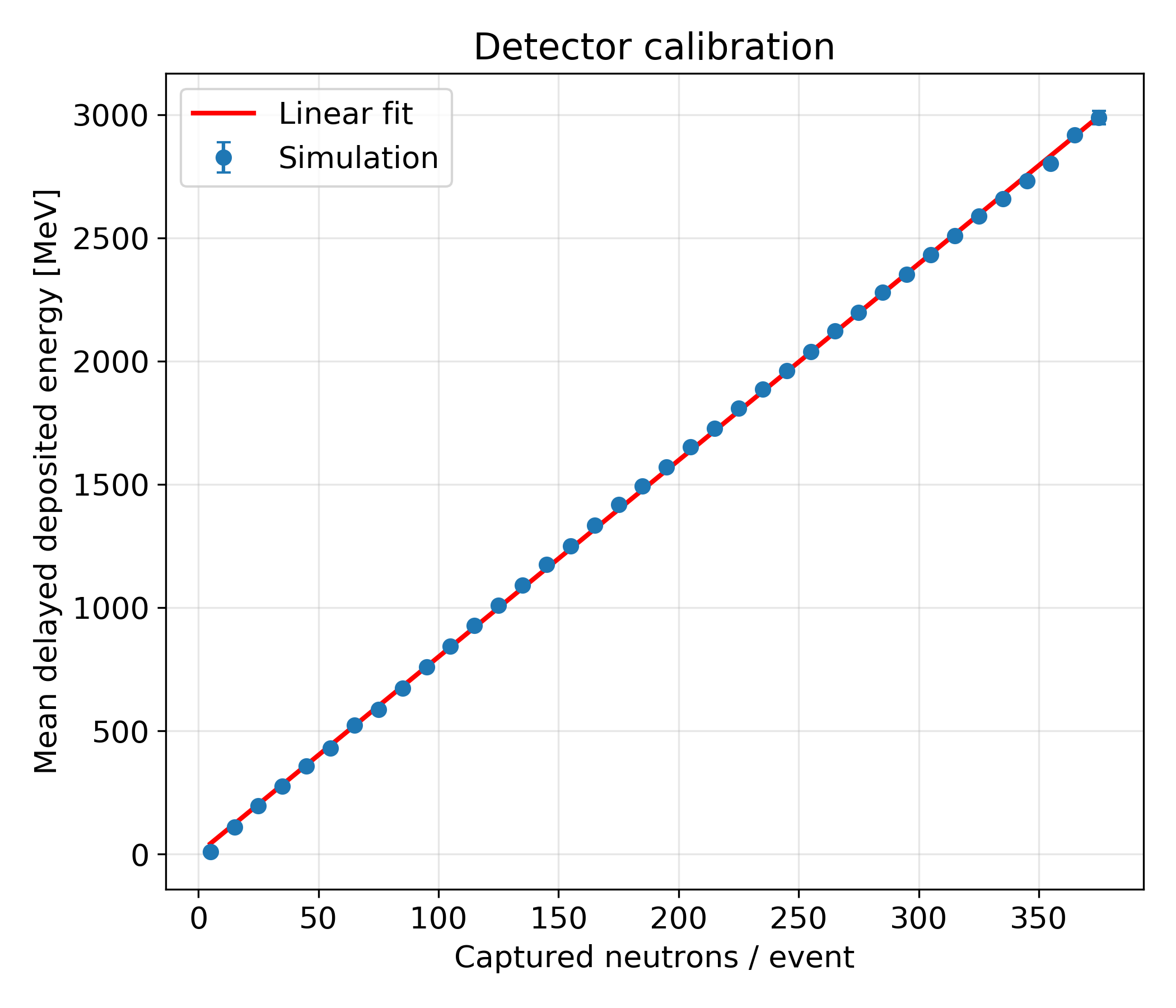}
\caption{
Left: event-by-event prompt deposited energy versus captured-neutron
multiplicity.
Center: delayed deposited energy versus captured-neutron multiplicity.
Right: calibration obtained from the delayed deposited energy as a
function of captured-neutron multiplicity.
}
\label{fig:prompt_delayed_corr}
\end{figure}

Figure~\ref{fig:prompt_delayed_corr} summarizes the three principal
event-by-event correlations obtained from the simulations.

The prompt deposited energy exhibits a broad correlation with the
captured-neutron multiplicity, although significant event-by-event
fluctuations are present. These fluctuations reflect the stochastic
nature of hadronic shower development and the varying fraction of
invisible energy associated with neutron production and nuclear breakup
processes~\cite{wigmans2000,Akchurin2009}. As discussed in the following
sections, the relationship is not purely linear, and its nonlinear
behaviour becomes evident when the prompt response is averaged in
narrow neutron-multiplicity intervals.

In contrast, the delayed deposited energy originating from neutron
captures is strongly correlated with the number of captured neutrons
($R^2=0.991$). This demonstrates that the delayed component provides a
quantitative observable associated with the neutron population produced
in the hadronic shower.

Plotting the delayed deposited energy as a function of the
captured-neutron multiplicity yields an almost ideal linear calibration
($R^2=0.9998$). The fitted slope corresponds to 7.97~MeV of deposited
delayed energy per captured neutron, consistent with the expected
$\sim8$~MeV total $\gamma$-cascade energy released following neutron
capture on gadolinium. This result provides a direct calibration of the
delayed signal under the simulated conditions.

\begin{figure}[htbp]
\centering
\includegraphics[width=0.31\textwidth]{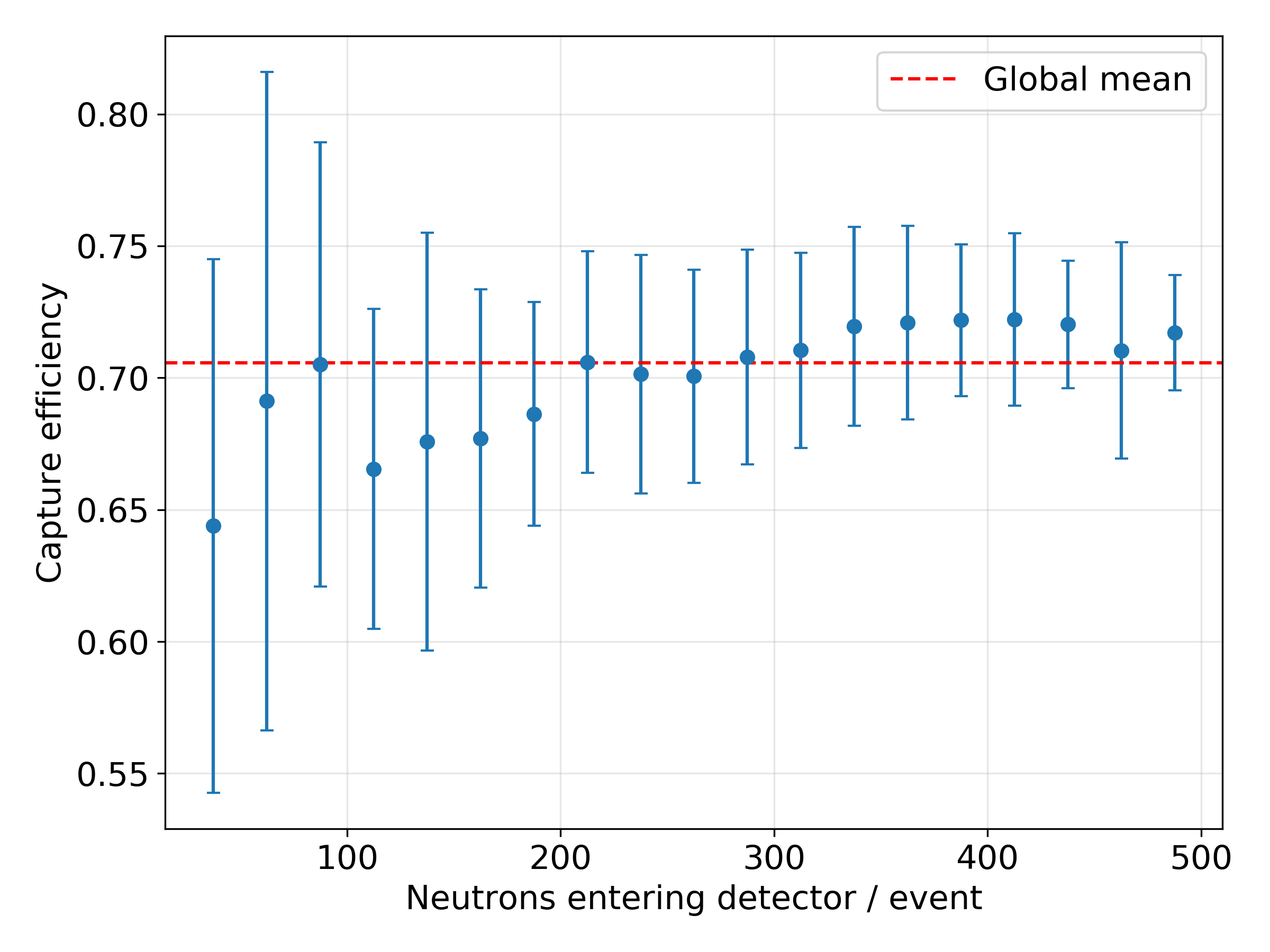}
\hfill
\includegraphics[width=0.31\textwidth]{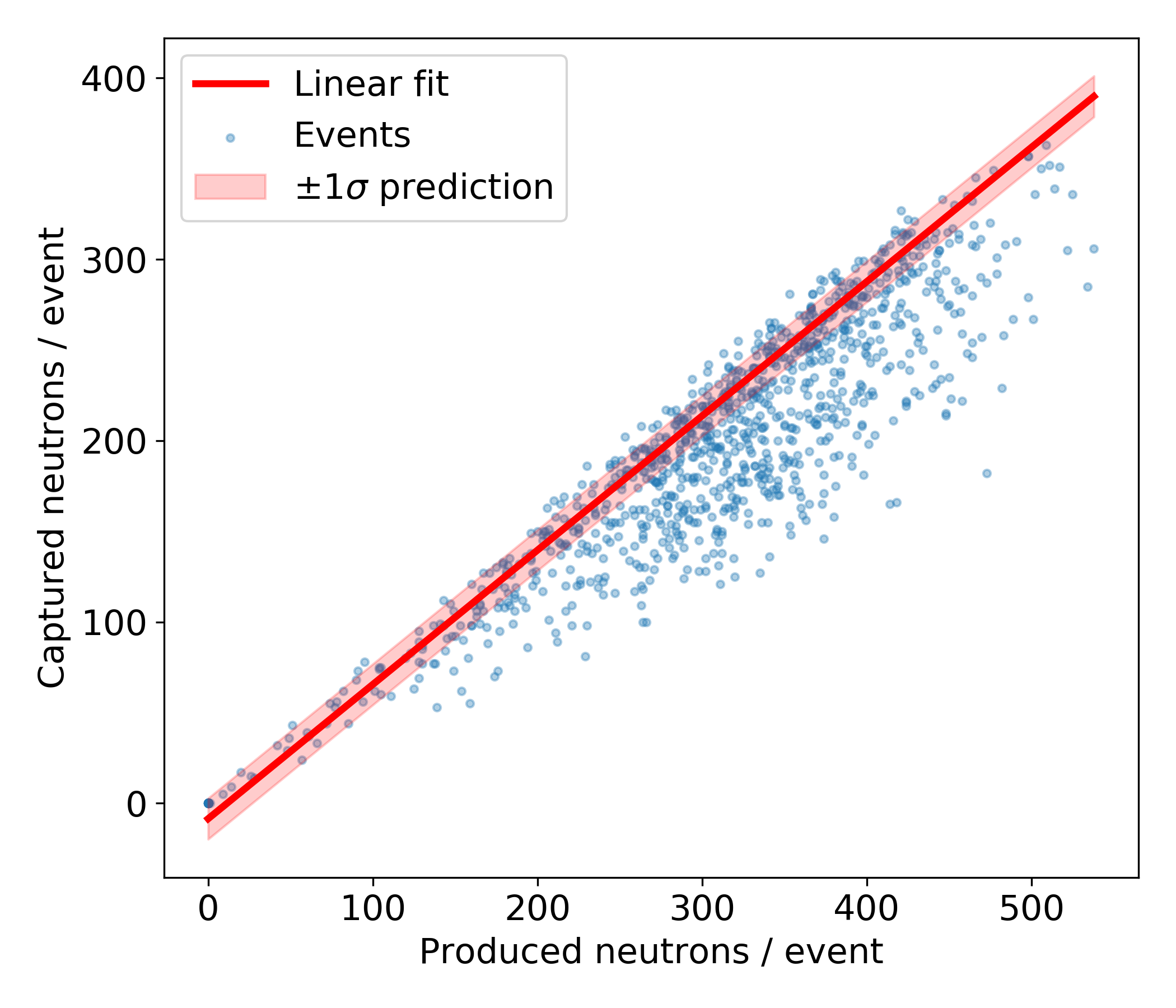}
\hfill
\includegraphics[width=0.31\textwidth]{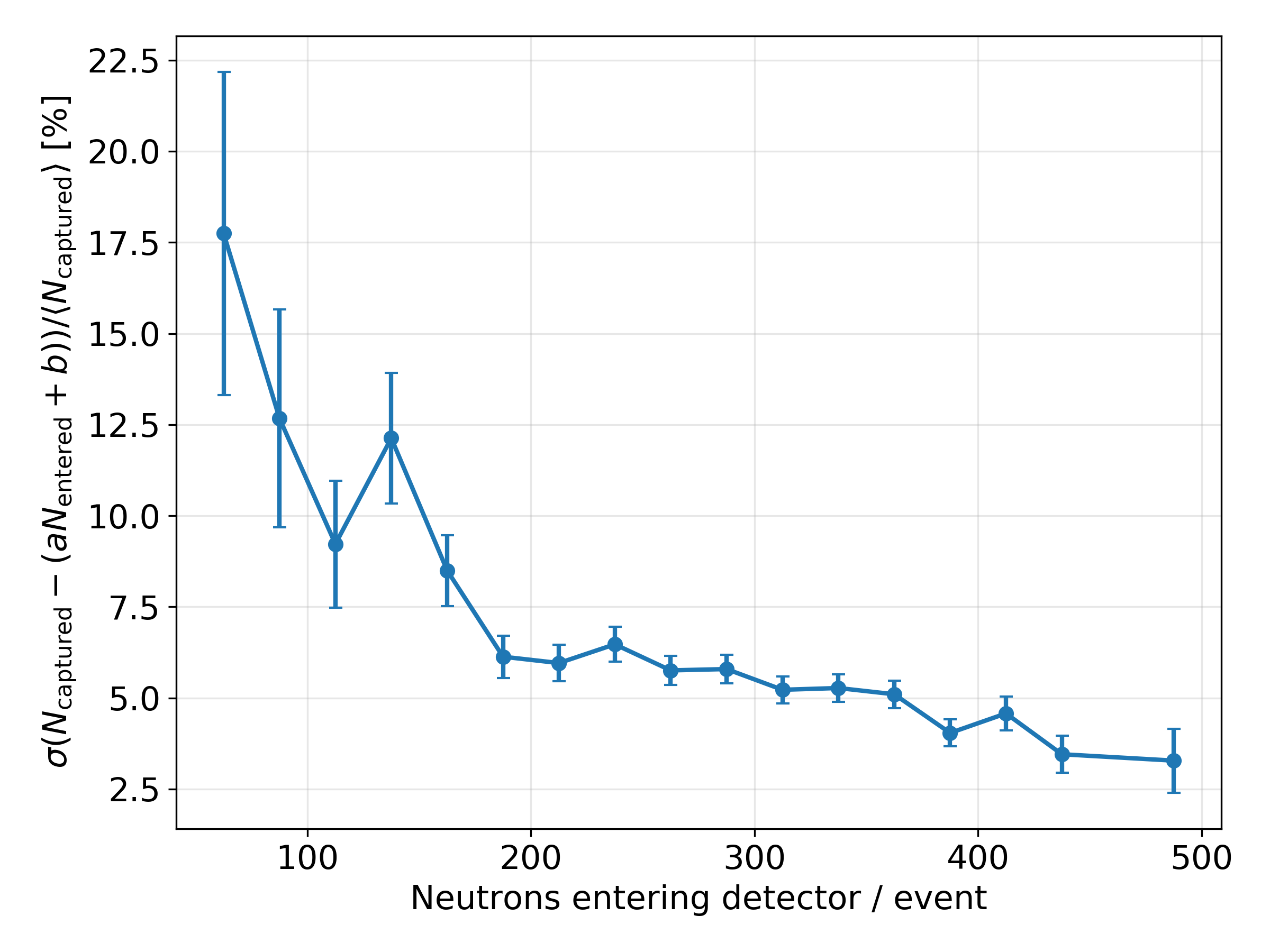}
\caption{
Neutron-counting performance of the six-layer calorimeter.
Left: neutron-capture efficiency versus the number of neutrons entering
the active volume.
Center: prediction band obtained from the event-by-event linear
calibration.
Right: neutron-counting resolution inferred from the relationship
between entering and captured neutron multiplicities. The results
characterize the simulated configuration.
}
\label{fig:performance}
\end{figure}

The practical neutron-counting performance was further evaluated using
the capture efficiency and the relationship between the number of
neutrons entering the active volume and the number finally captured.
Figure~\ref{fig:performance}a shows the neutron-capture efficiency as
a function of the entering neutron multiplicity. The efficiency
approaches a plateau around 70\% over the range relevant to the
present simulation.

Figure~\ref{fig:performance}b shows the 68\% prediction band associated
with the event-by-event linear calibration between entering and captured
neutron multiplicities. This band characterizes the spread around the
fitted relationship under the simulated conditions.
Figure~\ref{fig:performance}c presents the neutron-counting resolution
inferred from the relationship between entering and captured neutrons.
The approximately 7\% value obtained from this analysis characterizes
the neutron-counting fluctuations under the simulated conditions and
should not be interpreted as a general, standalone specification of
the detector resolution.
Taken together, these results indicate that the captured-neutron
multiplicity provides a useful measure of the neutron population
entering the active scintillator, although event-by-event fluctuations
remain. The strong correlation between the delayed signal and
neutron-capture multiplicity supports the use of delayed neutron
information as an additional observable for studying fluctuations in
the prompt calorimetric response.

The distinction between neutron-counting fluctuations and the overall
hadronic energy resolution is important. The approximately 7\%
neutron-counting resolution inferred here and the 21.8\% uncorrected
prompt-energy resolution describe different observables and cannot be
compared directly to quantify their respective contributions to the
overall calorimetric resolution. Nevertheless, the observed
correlations motivate investigating whether delayed neutron information
can be used to estimate part of the event-by-event variations in the
prompt response. The event-by-event correction examined in the
following sections explores this possibility~\cite{Akchurin2009}.
%%%%%%%%%%%%%%%%%%%%%%%%%%%%%%%%%%%%%%%%%%%%%%%%%%%%%%%%%%%%%%%%%%%%%%%%%%
\subsubsection{Correlation between prompt and delayed observables}
%%%%%%%%%%%%%%%%%%%%%%%%%%%%%%%%%%%%%%%%%%%%%%%%%%%%%%%%%%%%%%%%%%%%%%%%%%
\begin{figure}[htbp]
\centering
\includegraphics[width=0.47\textwidth]{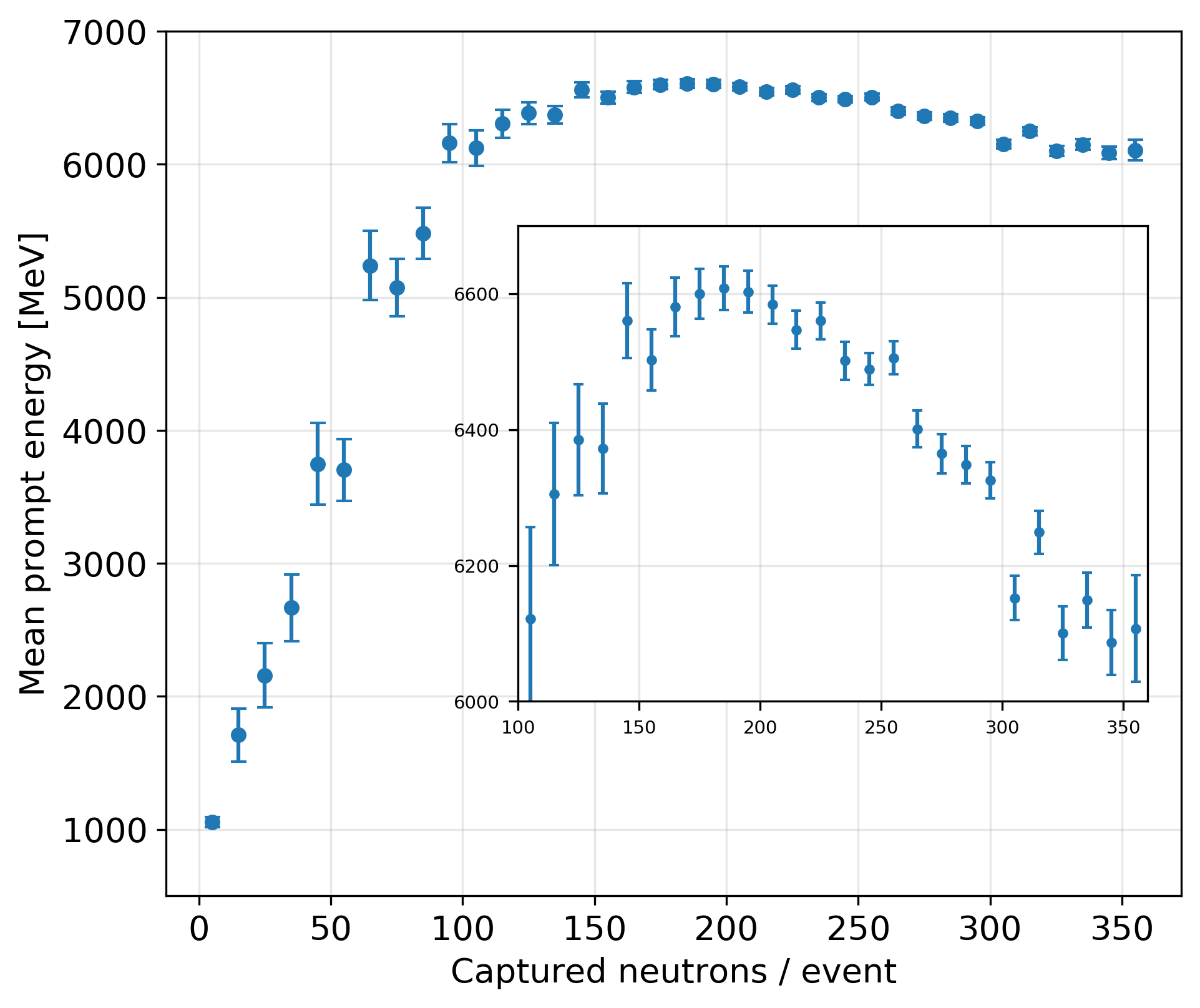}
\hfill
\includegraphics[width=0.47\textwidth]{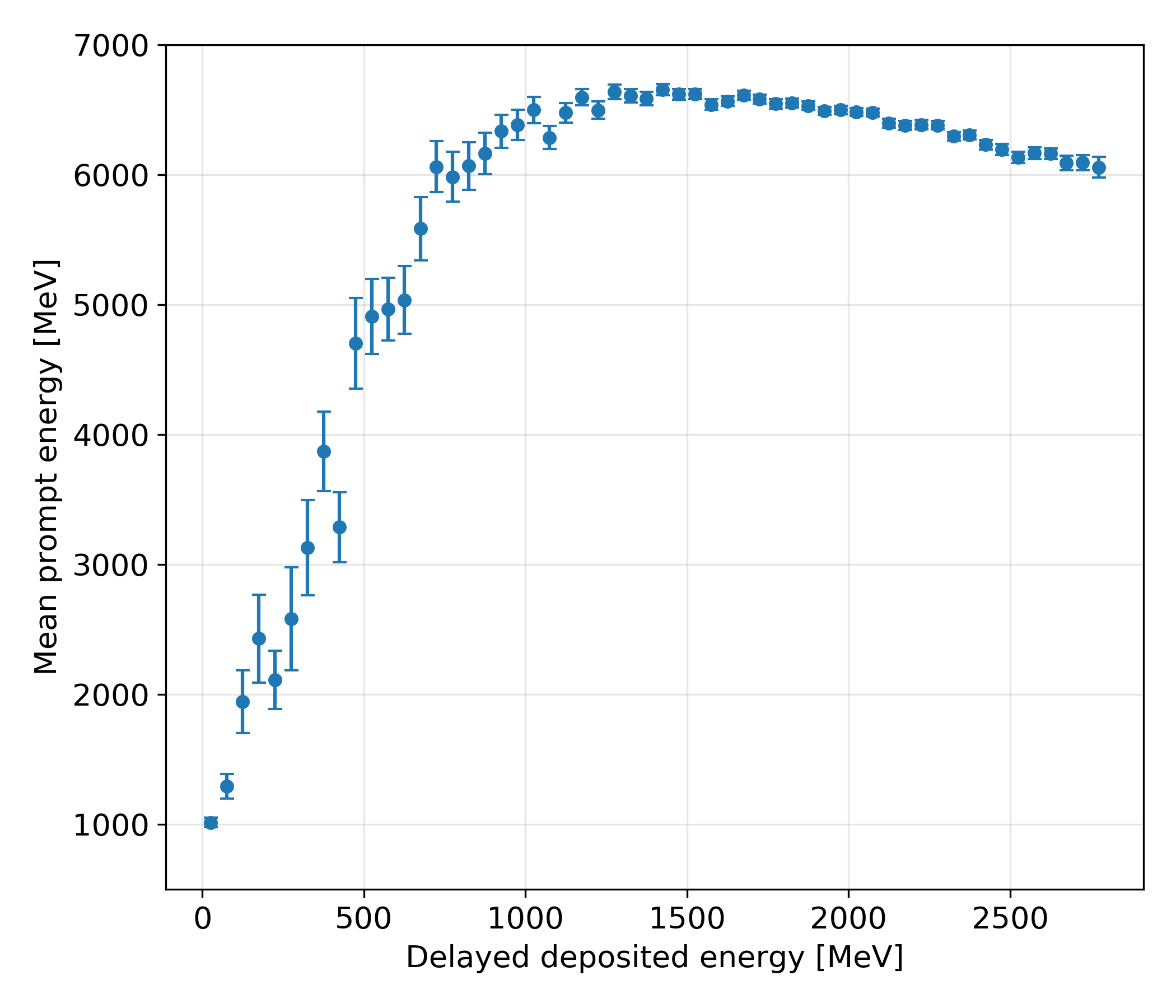}
\caption{
Average prompt deposited energy as a function of
(a) captured neutron multiplicity and
(b) delayed deposited energy.
The anti-correlation observed at large neutron multiplicities indicates that neutron production carries information on the invisible hadronic energy.
}
\label{fig:centroid}
\end{figure}
The delayed neutron signal was subsequently compared with the prompt calorimetric response recorded in the scintillator layers.

Figure~\ref{fig:centroid} presents the average prompt deposited energy as a function of the captured neutron multiplicity and, equivalently, as a function of the delayed deposited energy. To construct these curves, the event sample was divided into narrow intervals of captured-neutron multiplicity (or delayed deposited energy). For each interval, the corresponding prompt-energy distribution was projected and its centroid (mean value) was calculated. The resulting sequence of centroids defines the average prompt response as a function of the delayed neutron observable.

For small neutron multiplicities, the prompt deposited energy increases with the neutron yield. However, above approximately 180--200 captured neutrons the trend reverses and a clear anti-correlation appears.

This behaviour indicates that events producing unusually large numbers of evaporation neutrons deposit a smaller fraction of their energy promptly. Physically, a larger fraction of the primary proton energy is transferred to nuclear breakup and neutron production, reducing the immediately visible scintillation signal.

Since the delayed deposited energy is directly proportional to the neutron multiplicity, both observables carry essentially the same information. The delayed neutron signal therefore provides an experimental observable correlated with the invisible hadronic energy.
%%%%%%%%%%%%%%%%%%%%%%%%%%%%%%%%%%%%%%%%%%%%%%%%%%%%%%%%%%%%%%%%%%%%%%%%%%
\subsubsection{Effect of neutron multiplicity selection}
%%%%%%%%%%%%%%%%%%%%%%%%%%%%%%%%%%%%%%%%%%%%%%%%%%%%%%%%%%%%%%%%%%%%%%%%%%
\begin{figure}[htbp]
\centering
\includegraphics[width=0.60\textwidth]{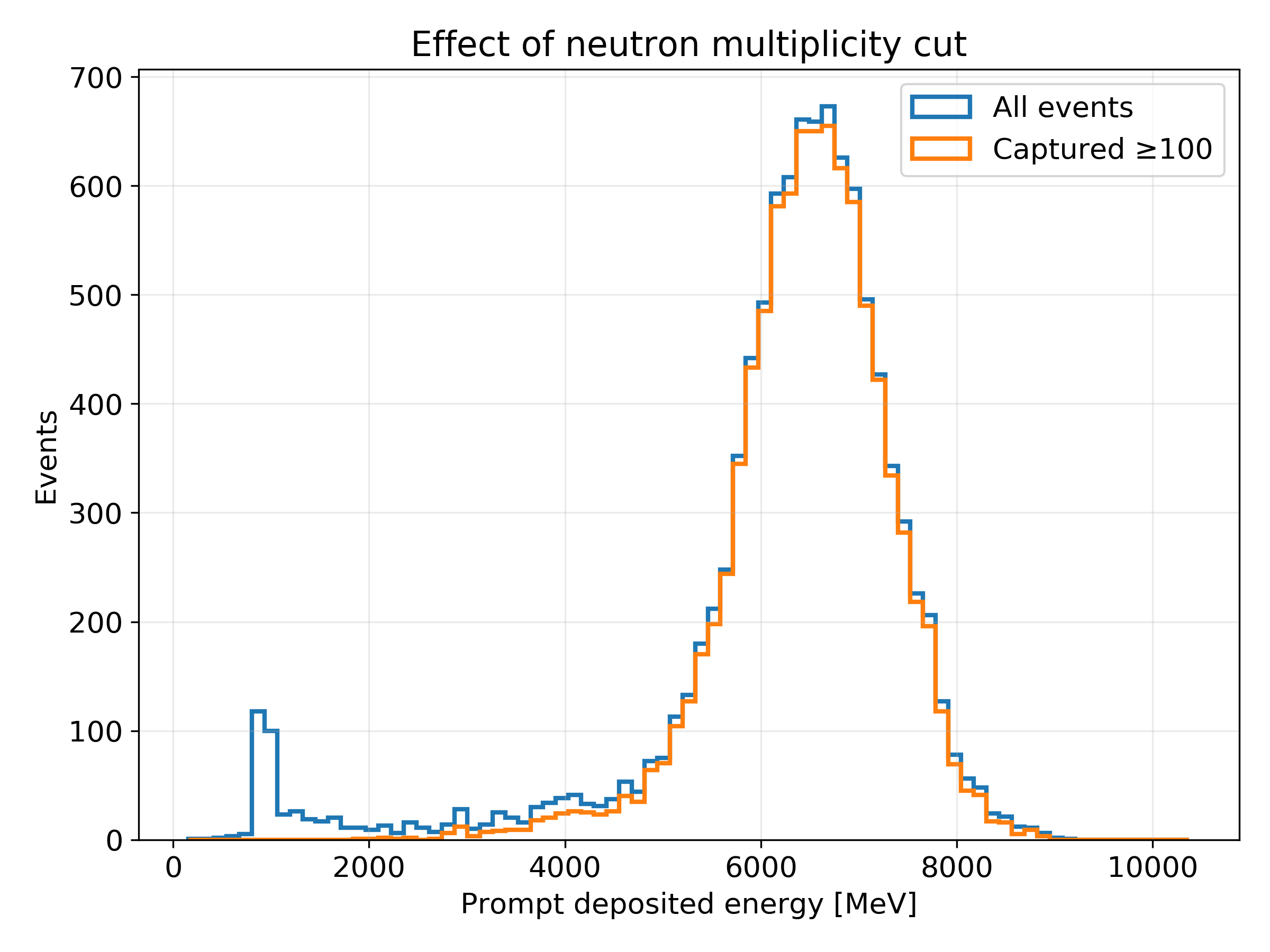}
\caption{
Prompt-energy spectrum for the complete event sample compared with the spectrum obtained after requiring more than 100 captured neutrons. The neutron multiplicity selection significantly narrows the energy distribution while retaining more than 91\% of the events.
}
\label{fig:cut100}
\end{figure}
The observed correlation between the prompt calorimetric response and the delayed neutron signal suggests that events producing larger neutron multiplicities may also exhibit reduced energy fluctuations. To illustrate this effect, progressively tighter lower thresholds were applied to the captured-neutron multiplicity, and the prompt-energy resolution was evaluated for the remaining events.

Figure~\ref{fig:cut100} compares the prompt-energy spectrum obtained from the complete event sample with that obtained after requiring more than 100 captured neutrons. A clear narrowing of the distribution is observed while retaining more than 91\% of the simulated events.

The evolution of the prompt-energy resolution for increasing neutron-multiplicity thresholds is summarized in Table~\ref{tab:MultiplicityCuts}. The apparent resolution continuously improves from 21.8\% for the complete event sample to approximately 6\% for the highest multiplicity selections.
\begin{table}[htbp]
\centering
\caption{Prompt-energy resolution obtained after applying progressively tighter lower thresholds on the captured-neutron multiplicity. The apparent improvement reflects the selection of increasingly homogeneous classes of hadronic showers rather than an event-by-event correction.}
\label{tab:MultiplicityCuts}

\begin{tabular}{cccc}
\hline
Minimum captured neutrons &
Events &
Mean energy (MeV) &
$\sigma/\mu$ (\%) \\
\hline
0   & 10000 & 6226 & 21.83 \\
100 & 9134  & 6479 & 12.64 \\
160 & 7332  & 6495 & 10.42 \\
200 & 5128  & 6450 &  9.21 \\
240 & 2806  & 6367 &  7.92 \\
280 & 1093  & 6237 &  6.80 \\
320 & 266   & 6093 &  6.14 \\
\hline
\end{tabular}
\end{table}
This behaviour demonstrates that events producing large neutron multiplicities constitute a more homogeneous class of hadronic showers with significantly smaller fluctuations in the prompt deposited energy. However, these values should not be interpreted as the intrinsic calorimeter resolution because the multiplicity cuts progressively reject an increasing fraction of the events. The cuts therefore represent an event selection rather than an event-by-event correction and cannot be applied in a realistic calorimetric measurement where the neutron multiplicity is not known a priori.

The significance of this study lies not in the absolute resolution
values themselves, but in the indication that fluctuations
associated with neutron production contribute to the event-by-event
variations in the prompt calorimetric response. This observation
motivates the investigation of event-by-event correction methods
based on the delayed neutron signal. Although the multiplicity
selection illustrates the relationship between neutron production
and prompt-energy fluctuations, it does not provide an event-by-event
correction and should not be interpreted as an improvement in the
intrinsic calorimeter resolution.
%%%%%%%%%%%%%%%%%%%%%%%%%%%%%%%%%%%%%%%%%%%%%%%%%%%%%%%%%%%%%%%%%%%%%%%%%%
\subsubsection{Prompt-energy resolution at fixed neutron multiplicity}
%%%%%%%%%%%%%%%%%%%%%%%%%%%%%%%%%%%%%%%%%%%%%%%%%%%%%%%%%%%%%%%%%%%%%%%%%%
\begin{figure}[htbp]
\centering
\includegraphics[width=0.47\textwidth]{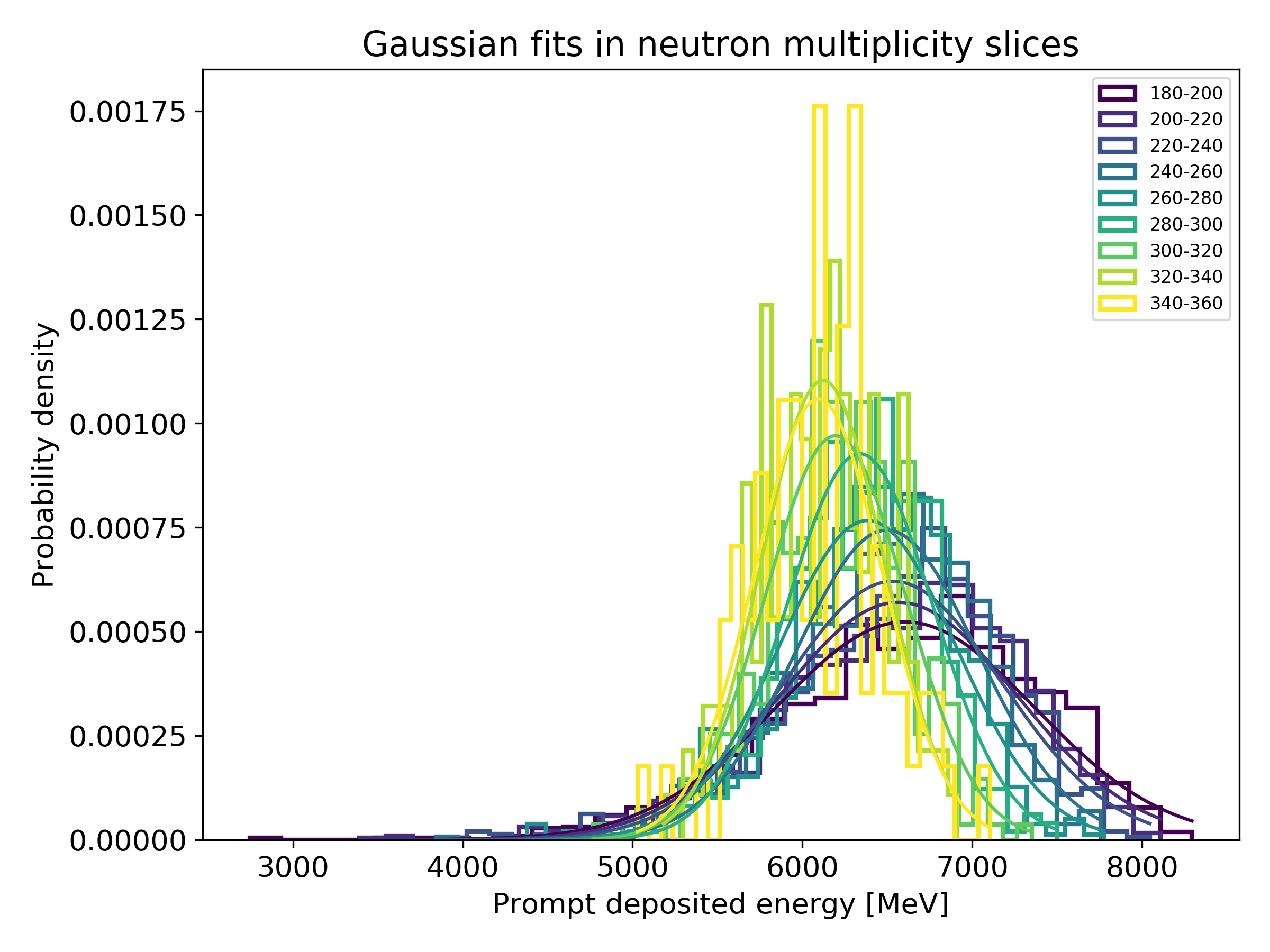}
\hfill
\includegraphics[width=0.47\textwidth]{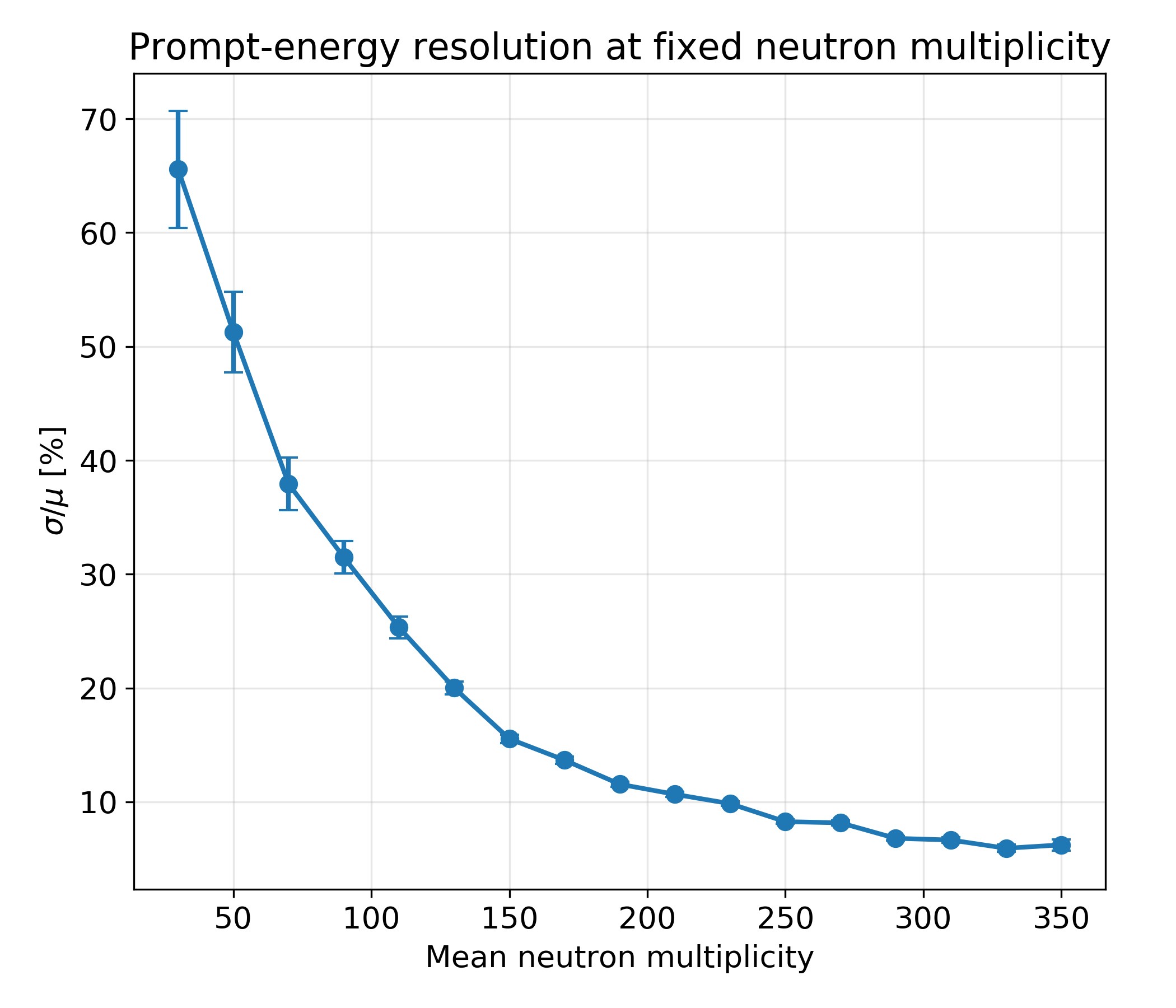}
\caption{
Prompt-energy resolution for fixed neutron multiplicity.
(a) Gaussian fits of representative neutron-multiplicity slices.
(b) Resolution as a function of the neutron multiplicity. Events
with similar neutron content exhibit narrower prompt-energy
distributions, indicating a smaller spread within each
multiplicity class.
}
\label{fig:slices}
\end{figure}

The previous analysis demonstrates that selecting events with large neutron multiplicities substantially improves the observed energy resolution. However, this does not explain the physical origin of the improvement.
To separate the intrinsic shower fluctuations from the fluctuations associated with neutron production, the prompt-energy resolution was evaluated inside narrow neutron-multiplicity intervals.
To investigate the origin of the calorimeter energy fluctuations, the events were grouped into narrow neutron-multiplicity intervals (20 captured neutrons wide). The prompt-energy resolution was then evaluated independently within each interval.

The results are summarized in Fig.~\ref{fig:slices}.

The prompt-energy resolution is found to improve systematically as the neutron multiplicity increases, decreasing from approximately 70\% for very low neutron multiplicities to about 6\% for the highest multiplicities. This behaviour demonstrates that the broad prompt-energy spectrum observed when all events are combined is not produced by a single class of hadronic showers. Instead, it results from the superposition of many event classes characterized by different neutron multiplicities, each exhibiting its own prompt-energy distribution and associated resolution.

The multiplicity-slice analysis shows that events with similar neutron
content exhibit narrower prompt-energy distributions than the complete
event sample. The width of the prompt-energy distribution therefore
depends strongly on the neutron multiplicity, and the broad inclusive
distribution reflects the superposition of event classes with
different neutron yields and different prompt responses.

This observation indicates that fluctuations associated with neutron
production are an important contribution to the overall variation in
the prompt calorimetric response. It does not, however, imply that all
prompt-energy fluctuations originate from neutron production, nor
that events with higher neutron multiplicities are intrinsically
better calorimetric events. Other fluctuations associated with hadronic
shower development and energy deposition remain.

The delayed neutron signal provides an observable correlated with
these variations and may therefore be used to correct part of the
event-by-event fluctuations in the prompt response. This motivates the
event-by-event correction investigated in the following section.
%%%%%%%%%%%%%%%%%%%%%%%%%%%%%%%%%%%%%%%%%%%%%%%%%%%%%%%%%%%%%%%%%%%%%%%%%%
\subsubsection{Event-by-event correction using the delayed neutron signal}
%%%%%%%%%%%%%%%%%%%%%%%%%%%%%%%%%%%%%%%%%%%%%%%%%%%%%%%%%%%%%%%%%%%%%%%%%%

\begin{figure}[htbp]
\centering
\includegraphics[width=0.60\textwidth]{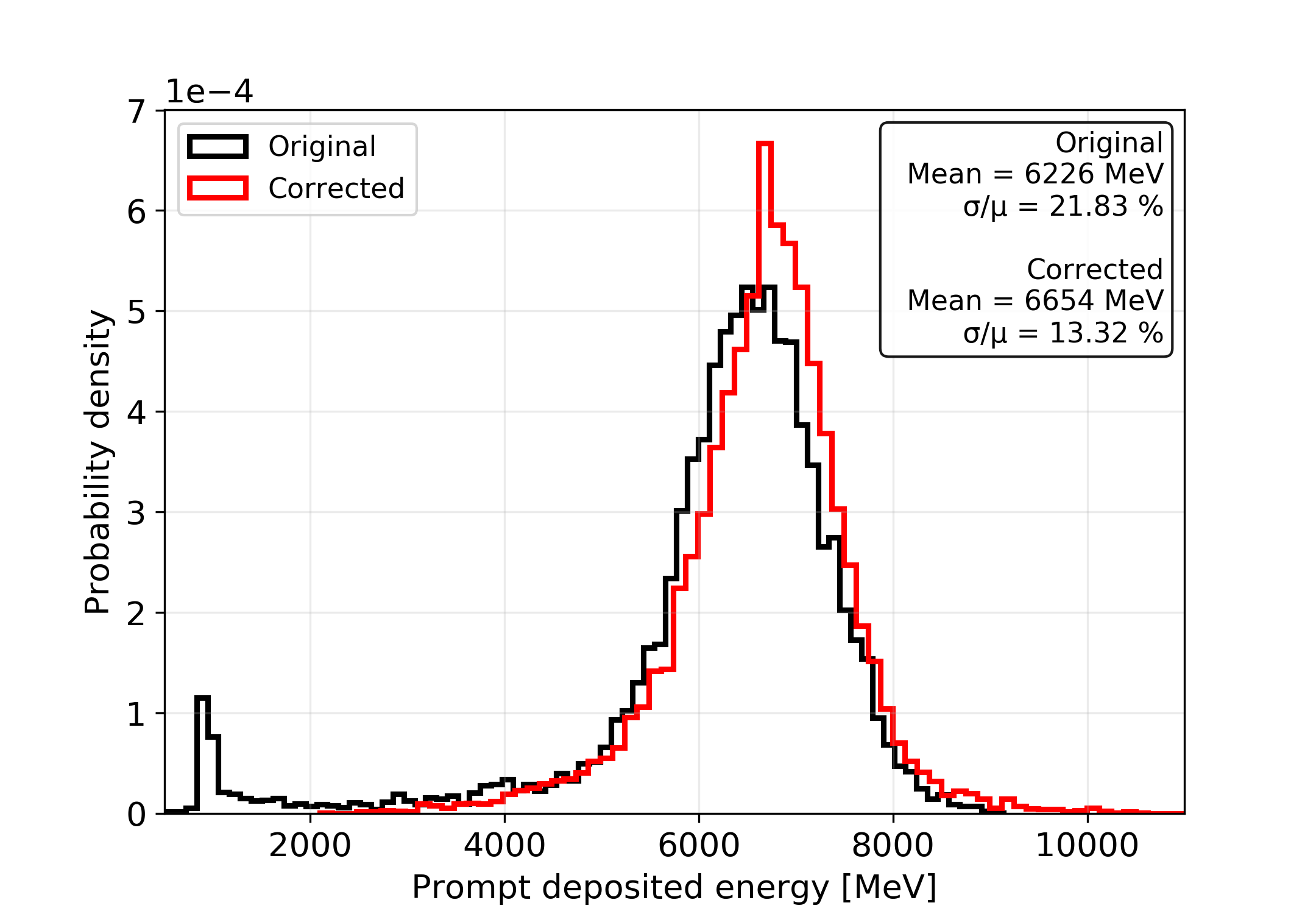}
\caption{
Prompt-energy spectrum before and after the event-by-event branch-wise interpolation correction based on the delayed neutron signal.
The correction improves the energy resolution from 21.8\% to 13.3\% without rejecting events.
}
\label{fig:correction}
\end{figure}

The detector characterization presented above demonstrates a strong
correlation between the delayed neutron signal and the neutron
population entering the active scintillator. The neutron-counting
performance analysis yielded an approximately 7\% resolution under
the simulated conditions. This value characterizes the fluctuations
in the relationship between entering and captured neutron
multiplicities and should not be interpreted as a general,
standalone specification of the detector resolution. The observed
correlations nevertheless motivate the use of the delayed neutron
signal as an additional observable for estimating variations in the
prompt calorimetric response.

Several correction algorithms were investigated, including linear and nonlinear neutron-based corrections. The best performance was achieved using a branch-wise interpolation of the average prompt energy as a function of the delayed deposited energy. The delayed-energy axis was divided into narrow intervals and, for each interval, the centroid of the prompt-energy distribution was determined. Connecting these centroids produced the average detector response curve shown in Fig.~\ref{fig:centroid}b. Since this curve exhibits two branches separated by a maximum, independent interpolation functions were constructed for the two regions. For each event, the measured delayed energy is used to estimate the average prompt response expected for that type of hadronic shower. The event is then shifted relative to this expected centroid, thereby compensating for fluctuations associated with the invisible hadronic energy while preserving the overall energy scale~\cite{Abramowicz1981,Lee2018}.

Figure~\ref{fig:correction} compares the original prompt-energy spectrum with the
corrected distribution. For the complete sample of 10\,000 simulated
10~GeV proton events, the event-by-event correction reduces the prompt
energy resolution from 21.8\% to 13.3\%, without rejecting events.
This corresponds to a substantial narrowing of the prompt-energy
distribution in the present simulation.

An important methodological limitation must be noted. The calibration
curve used to define the branch-wise interpolation functions and the
events used to evaluate the corrected energy resolution were obtained
from the same simulated sample. The reported improvement is therefore
an in-sample result and may overestimate the performance achievable
when the correction is applied to independent events. It should be
regarded as a preliminary demonstration of the potential of the
method, rather than as an independently validated resolution
improvement.

A robust assessment requires the calibration curve to be constructed
using a training sample and then applied, without refitting, to a
statistically independent validation sample. The original and
corrected prompt-energy resolutions should then be evaluated on the
same validation events. Such a procedure will establish whether the
observed improvement persists for events that were not used to derive
the correction.

The neutron-multiplicity selections discussed above provide a useful
complementary view of the relationship between neutron production and
prompt response, but they should not be confused with the
event-by-event correction. The selections change the composition of
the event sample by rejecting events, whereas the correction retains
all events and uses the delayed observable to estimate the average
prompt response associated with each event.

Despite this limitation, the present results demonstrate a clear
correlation between the delayed neutron observable and the prompt
calorimetric response. The branch-wise interpolation provides a
straightforward correction procedure that can be tested further using
independent event samples and more realistic calorimeter configurations.

%%%%%%%%%%%%%%%%%%%%%%%%%%%%%%%%%%%%%%%%%%%%%%%%%%%%%%%%%%%%%%%%%%%%%%
%%%%%%%%%%%%%%%%%%%%%%%%%%%%%%%%%%%%%%%%%%%%%%%%%%%%%%%%%%%%%%%%%%%%%%
\section{Discussion}
\begin{figure*}[htbp]
\centering
\includegraphics[width=0.94\textwidth]{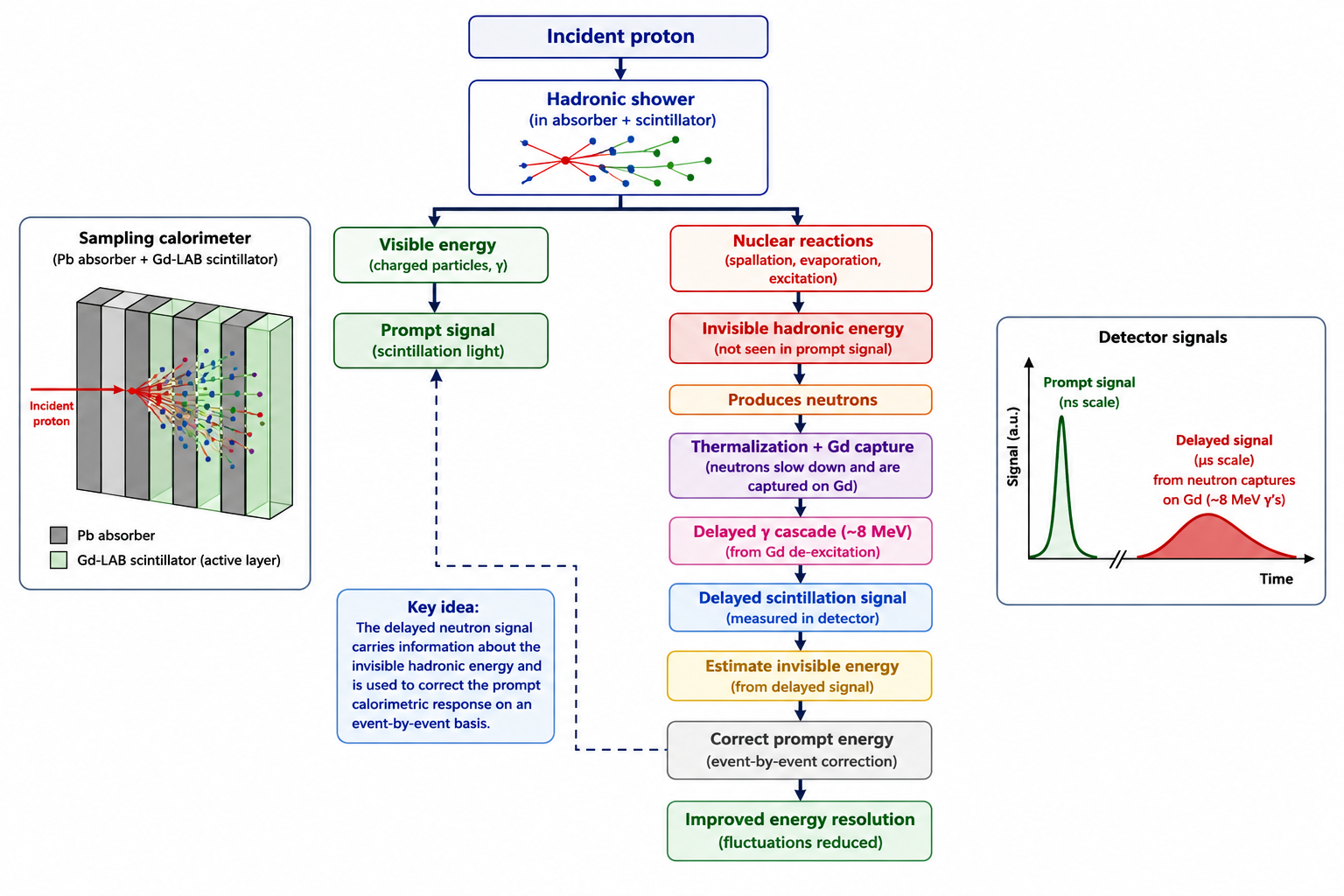}
\caption{
Physical interpretation of the neutron-sensitive calorimeter
concept and the event-by-event correction strategy investigated in
this work. Part of the hadronic shower energy appears promptly as
scintillation light, while nuclear reactions in the absorber produce
secondary neutrons. Following moderation and capture on gadolinium,
these neutrons generate delayed $\sim 8$~MeV $\gamma$ cascades and
a corresponding delayed scintillation signal. This signal provides
an additional observable correlated with the neutron component of
the shower and can be used to investigate possible corrections to
the prompt calorimetric response.
}
\label{fig:concept_summary}
\end{figure*}

%%%%%%%%%%%%%%%%%%%%%%%%%%%%%%%%%%%%%%%%%%%%%%%%%%%%%%%%%%%%%%%%%%%%%%
The analyses presented in this work demonstrate that the delayed neutron signal measured in Gd-loaded liquid scintillator layers carries substantial information on the invisible hadronic energy component of the shower. More importantly, this information can be exploited to improve the calorimetric energy reconstruction on an event-by-event basis.
Figure~\ref{fig:concept_summary} summarizes the physical picture emerging from the present study. A proton entering the calorimeter initiates a hadronic shower inside the lead absorber. Part of the incident energy is converted into ionization produced by charged particles and is immediately measured as the prompt scintillation signal. The remaining fraction is consumed in nuclear reactions inside the absorber, including spallation, nuclear excitation and evaporation processes, producing secondary neutrons and other non-visible products. This energy is commonly referred to as the invisible hadronic energy because it does not contribute directly to the prompt calorimetric response.
The secondary neutrons subsequently undergo moderation inside the
detector and are captured predominantly on gadolinium nuclei. Each
capture releases a characteristic $\sim 8$~MeV $\gamma$ cascade,
which produces a delayed scintillation signal. The delayed deposited
energy is strongly correlated with the neutron-capture multiplicity
and therefore provides an estimate of the neutron component of
each hadronic shower.

The event-by-event analyses show that the neutron content is strongly
correlated with variations in the prompt calorimetric response.
Events with different neutron multiplicities exhibit different
prompt-energy distributions, whereas events with similar neutron
multiplicities tend to have narrower prompt-energy distributions.
This indicates that fluctuations associated with neutron production
represent an important contribution to the overall variations in
the prompt response. However, the present analysis does not establish
that neutron production accounts for all, or necessarily the largest
part, of the hadronic energy fluctuations.

The prompt-energy centroid as a function of neutron multiplicity
also reveals an anti-correlation at high multiplicities. Events
producing more neutrons tend to deposit slightly less energy
promptly, consistent with a larger fraction of the shower energy
being transferred to nuclear processes and neutron production
rather than to prompt ionization. A similar behaviour is observed
when the delayed deposited energy is used instead of the neutron
multiplicity. This supports the interpretation of the delayed
signal as an observable correlated with the invisible component
of the hadronic shower.
Based on these correlations, the delayed neutron signal was used
to estimate the average prompt response associated with each
event. The correction giving the best performance in the present
study was based on a branch-wise interpolation of the prompt-energy
centroid as a function of delayed deposited energy. Rather than
applying a single global linear correction, this procedure uses
the measured delayed energy to estimate the expected prompt response
and adjusts the event accordingly.

For the complete sample of 10\,000 simulated 10~GeV proton events,
the branch-wise correction reduces the prompt-energy resolution
from 21.8\% to 13.3\%, without rejecting any events. This corresponds
to a reduction of approximately 63\% in the variance of the
prompt-energy distribution in the present sample. In addition,
the neutron-multiplicity slice analysis shows that events with
similar neutron content exhibit narrower prompt-energy
distributions, with resolutions approaching 6\% in the highest
multiplicity intervals. These observations indicate that
neutron-production fluctuations contribute significantly to
the calorimetric response variations. However, the correction
was evaluated using the same sample from which its calibration
was derived, and its performance must therefore be assessed
using independent events before drawing conclusions about the
achievable improvement in calorimetric resolution.

The observed dependence of the prompt-energy resolution on neutron
multiplicity also motivates studies at different incident-particle
energies. Since the neutron yield and the development of hadronic
showers vary with incident energy, it remains to be established
whether the correlations observed here persist over a broader
energy range and whether the correction retains its effectiveness.

The event-by-event correction presented here should nevertheless be
interpreted in the context of the present simulation. In particular,
the calibration and evaluation were performed using the same sample
of simulated proton events. Independent validation is therefore
required to quantify the performance of the correction on previously
unseen events and to establish its robustness beyond the specific
configuration studied here.

The concept may also be relevant to the broader development of
high-performance hadronic calorimetry for future collider experiments.
At the Future Circular Collider in electron--positron mode (FCC-ee),
precision reconstruction of hadronic energies and masses is an
important detector-performance goal, particularly in the context of
particle-flow-based event reconstruction~\cite{Aleksa2021FCCee}.
The approach investigated in this work provides an additional
observable associated with neutron production and capture that could
complement the conventional prompt calorimetric signal.

The present study, however, is based on single 10~GeV proton
interactions and a preliminary six-layer Pb/LAB-Gd geometry. It does
not establish an improvement in jet-energy or jet-mass resolution,
nor does it demonstrate performance in a realistic FCC-ee detector
environment. These questions require dedicated simulations of
multi-particle hadronic showers, including realistic jet fragmentation
and an assessment of the detector response.

A natural extension of this work is therefore to investigate the
proposed concept using hadronic jets over a representative energy
range. Such studies could determine whether the correlations observed
for single protons remain useful when the calorimeter is exposed to
the superposition of showers initiated by the particles within a jet.
They should also assess the effects of shower overlap, energy leakage,
the integration window used to collect the delayed signal, and the
implementation of the neutron-sensitive layers in a realistic
calorimeter design.

Together with independent validation of the event-by-event correction,
these investigations will be necessary to determine whether the
observed improvement in prompt-energy resolution can be maintained
under more realistic calorimetric conditions and whether delayed
neutron information can contribute to jet-energy reconstruction.

To provide an experimental validation of the proposed detection concept, a staged proof-of-concept programme could be pursued, beginning with a small-scale demonstrator. An initial single-cell setup with dimensions of $5 \times 5 \times 5~\text{cm}^3$ could be used to investigate the detection of delayed signals associated with neutron capture on gadolinium in a LAB-based scintillator. Such a setup could employ readily available components, including a Gd-loaded LAB scintillator, an optical enclosure, and suitable photosensors, such as avalanche photodiodes (APDs), subject to the requirements of the optical readout and timing system. The initial measurements would aim to characterize the delayed capture signal and distinguish it from the prompt scintillation component. Following this first stage, a modest $3 \times 3$ matrix prototype could be developed to investigate cell-to-cell response variations, spatial effects, and correlations between prompt and delayed signals. This staged approach would provide an experimental basis for assessing the feasibility of the concept before proceeding to more complex prototype configurations or test-beam studies.
%%%%%%%%%%%%%%%%%%%%%%%%%%%%%%%%%%%%%%%%%%%%%%%%%%%%%%%%%%%%%%%%%%%%%%
%%%%%%%%%%%%%%%%%%%%%%%%%%%%%%%%%%%%%%%%%%%%%%%%%%%%%%%%%%%%%%%%%%%%%%
\section{Conclusions}

This study investigates the use of delayed neutron-capture signals
in a Gd-loaded Pb/LAB sampling calorimeter as an additional
observable for characterizing hadronic shower fluctuations.
The Geant4 simulations show a strong correlation between the
delayed deposited energy and the captured-neutron multiplicity.
The corresponding calibration has a slope consistent with the
expected energy released in the $\gamma$-cascade following
neutron capture on gadolinium.

The simulated six-layer configuration captures approximately 71\% of the neutrons entering the active scintillator volume, while the delayed response provides information complementary to the prompt calorimetric signal. The prompt-energy distributions observed in narrow neutron-multiplicity intervals indicate that fluctuations associated with neutron production are an important component of the overall response variations.

A preliminary event-by-event correction based on the delayed-energy observable reduces the prompt-energy resolution from 21.8\% to 13.3\% for the simulated 10~GeV proton sample. Since the correction was developed and evaluated using the same sample, this improvement should be regarded as an encouraging preliminary result requiring independent validation.

Further studies should assess the method using independent simulated samples, more realistic detector and readout models, and multi-particle hadronic showers and jets. In parallel, a staged experimental programme, beginning with a single-cell demonstrator and progressing towards a small matrix prototype, could investigate the feasibility of measuring the delayed neutron-capture signal and exploiting its correlation with the prompt response.

These results establish a simulation-based proof of concept for using delayed neutron information to investigate and potentially correct fluctuations in hadronic calorimetry. They motivate further simulation and experimental studies to determine the robustness and practical applicability of the proposed approach.

\bibliographystyle{IEEEtran}
\bibliography{refs}

\end{document}